\documentclass[onecolumn]{emulateapj}
\usepackage{apjfonts}

\newcommand{\bomega}{\mbox{\boldmath$\omega$}}
\newcommand{\bnabla}{\mbox{\boldmath$\nabla$}}
\newcommand{\Om}{\mbox{\boldmath$\Omega$}}
\newcommand{\bmu}{\mbox{\boldmath$\mu$}}

\newcommand{\be}{\begin{equation}}
\newcommand{\ee}{\end{equation}}
\newcommand{\beq}{\begin{eqnarray}}
\newcommand{\eeq}{\end{eqnarray}}
\newcommand{\nn}{\mbox{} \nonumber \\ \mbox{} }
\newcommand{\ba}{\begin{eqnarray}}
\newcommand{\ea}{\end{eqnarray}}

\newcommand{\B}{{\bf B}}

\newcommand\lo{\mathrel{\raise.3ex\hbox{$<$}\mkern-14mu\lower0.6ex\hbox{$\sim$}}}
\newcommand\go{\mathrel{\raise.3ex\hbox{$>$}\mkern-14mu\lower0.6ex\hbox{$\sim$}}}
\def\simlt{\lower.5ex\hbox{$\; \buildrel < \over \sim \;$}}
\def\simgt{\lower.5ex\hbox{$\; \buildrel > \over \sim \;$}}
\def\BQ{B_{\rm Q}}

\def\bB{{\mathbf B}}
\def\RNS{R_{\rm NS}}
\def\BNS{B_{\rm NS}}
\def\MNS{M_{\rm NS}}
\def\INS{I_{\rm NS}}
\def\K3{\widetilde \kappa_3}
\def\eps3{\widetilde \varepsilon_3}

\def\rhoGJ{\rho_{\rm GJ}}
\def\IGJ{I_{\rm GJ}}
\def\kapNS{\kappa_{\rm NS}}

\def\Tbb{T_{\rm bb}}
\def\kB{k_{\rm B}}

\def\lbar{\rlap{$\lambda$}^{\_\_}_{\;\;\;e}}
\def\Mgam{{\cal M}_\gamma}
\def\Mpm{{\cal M}_\pm}

\def\Epar{E_\parallel}

\def\gmax{\gamma_{\rm max}}

\def\thkB{\theta_{kB}}
 \def\mukB{\mu}
\def\mumin{\mu_{\rm min}}

\def\Th{\Theta_{\rm bb}}
\def\Thh{\left({k_{\rm B}T_{\rm bb}\over 0.5~{\rm keV}}\right)}

\def\ndot{\dot{n}_\pm}
\def\out{\rhd}
\def\in{\lhd}
\def\Gresout{\Gamma^{\rm res}_\out}
\def\Gresin{\Gamma^{\rm res}_\in}
\def\yout{\widetilde\omega_\out}
\def\yin{\widetilde\omega_\in}

\def\gamout{\gamma_\out}
\def\gamin{\gamma_\in}
\def\nout{n_\out}
\def\nin{n_\in}
\def\rhoout{\rho_\out}
\def\rhoin{\rho_\in}
\def\Iout{I_\out}
\def\Iin{I_\in}

\def\alf{\alpha_f}

\shorttitle{ELECTRODYNAMICS OF MAGNETARS:  INNER ACCELERATOR}
\shortauthors{THOMPSON}
\slugcomment{Submitted to the Astrophysical Journal}
\begin{document}
\title{ELECTRODYNAMICS OF MAGNETARS IV:\\
       SELF-CONSISTENT MODEL OF THE INNER ACCELERATOR\\
       WITH IMPLICATIONS FOR PULSED RADIO EMISSION}

\author{Christopher Thompson}
\affil{CITA, 60 St. George St., Toronto, ON M5S 3H8, Canada}

\begin{abstract}
  We consider the voltage structure in
  the open-field circuit and outer magnetosphere of a magnetar.  
  The standard polar-cap model for radio pulsars is modified
  significantly when the polar magnetic field exceeds $1.8\times 10^{14}$ G.
  Pairs are created by accelerated particles via resonant scattering of thermal
  X-rays, followed by the nearly instantaneous conversion of the 
  scattered photon to a pair.  A surface gap is then efficiently screened
  by $e^\pm$ creation, which regulates the voltage in the inner part of the
  circuit to $\simlt 10^9$~V.  We also examine the electrostatic gap structure
  that can form when the magnetic field is somewhat weaker, and deduce a
  voltage 10-30 times larger over a range of surface temperatures.  
  We examine carefully how the flow of charge back to the star above the gap
  depends on the magnitude of the current that is 
  extracted from the surface of the star, on the curvature of the
  magnetic field lines, and on resonant drag.  The rates of different
  channels of pair creation are determined self-consistently, including the
  non-resonant scattering of X-rays, and collisions between gamma rays and
  X-rays.  We find that the electrostatic gap solution has too small
  a voltage to sustain the observed pulsed radio output of magnetars 
  unless i) the magnetic axis is nearly aligned with the rotation axis 
  and the light of sight; or ii) the gap is present on the closed as well as
  the open magnetic field lines.  Several properties of the radio magnetars --
  their rapid variability, broad pulses, and unusually hard radio spectra --
  are consistent with a third possibility, that the current in the outer
  magnetosphere is strongly variable, and a very high rate of pair creation
  is sustained by a turbulent cascade.  

\end{abstract}


\section{Introduction}\label{s:one}

Magnetars are persistent sources of broad band, non-thermal radiation
(see Woods \& Thompson 2006 for a recent review).
The discovery of pulsed radio emission from XTE J1810$-$197
and 1E 1547.0$-$5408 (Halpern et al. 2005; Camilo et al. 2006, 
2007a,b) adds to the body of evidence identifying them as isolated,
non-accreting neutron stars.  This radio emission is peculiar in
several respects:  its
brightness fluctuates strongly on a timescale of minutes to days, 
and peaks above a frequency of 100 GHz.  The detection of
pulsed radio emission from two magnetars, some 15 percent of the
presently known population, appears to
be a consequence of the surprisingly large pulse duty cycle.
The known magnetars spin with typical periods of several seconds,
and their open field lines form a narrow bundle close to the star.
For this reason, it had generally been expected that
their radio emission would be strongly beamed.

The output of a magnetar in the hard X-ray band can
greatly exceed the spindown power of the star (Kuiper et al. 2004, 2006), and 
so most of the underlying dissipation is probably concentrated in the 
closed magnetosphere (Thompson \& Beloborodov 2005; Baring \& Harding
2007).  
The same energetic argument does not apply to the radio output, 
which is smaller than the spindown luminosity.
The radio emission zone could, therefore, be restricted to the open magnetic
field lines.  

It has also been suggested that relatively strong currents flow within the
outer magnetosphere of a magnetar, being sustained by the outward transfer
of twist from an inner zone where it is injected by starquakes.
This redistribution of the twist causes
the poloidal field lines to expand slightly, thereby increasing
the open-field voltage (Thompson et al. 2002).   The effect depends 
on the extraction of
a tiny fraction the helicity that is stored in the inner magnetosphere
and stellar interior.  The radio brightness of
the magnetar XTE J1810$-$197 is observed to correlate strongly with
the spindown torque (Camilo et al. 2007a), 
a behavior which is consistent with this type of
structural change.  Indeed, the torque variations measured in
XTE J1810$-$197 are comparable in magnitude to those seen in the 
Soft Gamma Repeaters, which are much more variable in the X-ray band
(Woods et al. 2002).  

The injection of twist into the outer magnetosphere
inevitably triggers current-driven instabilities and strong 
fluctuations in the current.  
A detailed model of pair creation in a dynamic open
magnetic flux tube is given in a companion paper (Thompson 2008; 
see also Lyutikov \& Thompson 2008).
The torsional oscillations of the magnetic field
cascade to high frequencies if their amplitude exceeds a critical
value, resulting in a very high rate of pair creation.

In this paper, focus on the case where the current that is drawn
from the magnetar surface varies only slowly on the timescale $\RNS/c$, where
$\RNS$ is the stellar radius.  Our main goal is to work out the structure of a 
pair-creating diode forming at the surface when the magnetic field
is stronger than the QED value $\BQ = 4.4\times 10^{13}$ G.
 The voltage across the diode is influenced
by several competing effects:  the mechanism(s) of pair creation;
the charge flow in the outer magnetosphere;
and the gradient in the corotation charge density $\rhoGJ$ at the stellar
surface.   Our solution applies, in principle, to both the open and
the extended closed magnetic field lines, because it depends only
on the flux of charges returning from the outer magnetosphere.

A detailed review of the basic QED processes leading to pair
creation in super-QED magnetic fields is given in Usov \&
Melrose (1996), and a companion paper (Thompson 2008).  The dominant
mechanism of pair creation involves the creation of gamma rays
by resonant scattering
of thermal X-rays (Kardashev et al. 1984, Daugherty \& Harding 1989, 
Sturner 1995, and Hibschman \& Arons 2001a,b).  
The voltage across a gap can also be regulated by collisions
between gamma rays and thermal X-rays (e.g. Zhang \& Qiao 1998) and,
as we show in this paper, by the direct conversion of non-resonantly
scattered X-rays to pairs.

One of our main results is that a diode {\it cannot} be self-consistently
be maintained at the surface when the magnetic field $\BNS
 > 4\BQ = 1.8\times 10^{14}$ G.  In this case,
the scattered photon is typically
above threshold for pair creation, and there is essentially no
delay between the creation of a gamma ray 
and its conversion to a pair  (Beloborodov \& Thompson 2007).
We describe a circuit solution that
has an inward $e^\pm$ discharge, within which the voltage that is
screened to a low value $\simlt 10^9$~V.  The stability of this 
solution is also investigated,
and is shown explicitly to depend on a delay between gamma-ray emission 
and pair creation.  

A surface gap is possible when $\BNS < 4\BQ$ and the current density
is larger than $\rhoGJ c$.  We construct a detailed
model including all the relevant modes of pair creation. 
We focus on space-charge limited flows, and explain why charges of
both signs can be supplied continuously from the surface of a magnetar.
Even through heavy atoms are tightly bound into molecular chains
(Medin \& Lai 2007, and references therein), ions can be liberated
from the magnetar surface by photoionization and knockout processes.

The equilibrium gap voltage is calculated as a function of the X-ray
black body temperature $\Tbb$ of the stellar surface.  It is gradually
reduced as $\kB\Tbb$ rises above $\sim 0.15-0.2$ keV, and
non-resonant scattering and photon collisions supply increasing
amounts of pair creation within the gap. 
 Pairs that are created just outside the
gap continue to lose energy by resonant scattering, and 
the resulting increase in the pair multiplicity is derived.  We
explain in detail how the polarization of a pair-creating cloud
that is formed just outside the diode will create enough returning
charges to compensate the gradient in corotation charge density
outside the gap.  

An important subtlety involves the effect of
cyclotron drag at $\sim 10$ stellar radii;  we show that outside
this zone the correct space charge is obtained if the charges
of sign opposite to $\rhoGJ$ flow outward sub-relativistically.
This solution breaks down beyond a critical angle from the magnetic
axis, where charges of both signs can be decelerated to trans-relativistic
speeds.  As a result, we point out that there is a selection effect
against observing pulsed radio emission from X-ray bright neutron stars
with spin periods shorter than $\sim 0.3$ seconds.  The reflection
of gamma rays back toward the star via resonant scattering can have
a similar effect at much longer spin periods.

The plan of the paper is as follows.  
The origin of the open-field voltage is reviewed in \S~\ref{s:two},
and the two basic types of circuit solution -- with and without a diode --
are described. 
The voltage of a surface diode in a magnetic field $\la 4\BQ$ is derived
in \S \ref{s:four}, and the effects of cyclotron drag are investigated
in \S \ref{s:cycdrag}.  In section 
\ref{s:five} we explore in detail the plasma flow on the open field 
lines when $B_{\rm NS} > 4\BQ$ and a diode is absent at the surface
of the star.    The paper closes with a summary of how
the different circuit solutions described in the paper are constrained
by the luminosities of the radio magnetars.   The Appendix 
compares the effects of field-line bending and general-relativistic 
frame dragging on the gradient in corotation charge density.

\section{Voltage Generation}
\label{s:two}

In the absence of pair creation, a very large voltage develops
on the open magnetic field lines of a rotating neutron star.
When the effects of screening by pair creation
are included, the magnitude and time dependence of the voltage are believed
to depend on several factors
(e.g. Sturrock 1971; Ruderman \& Sutherland 1975; Scharlemann, Arons, 
\& Fawley 1978).    A vaccum gap could form with a relatively large 
and variable voltage when the corotation charge density is positive,
if the ions are tightly bound in molecular chains to the neutron star surface.
Although magnetars are indeed expected to have condensed surfaces
in the absence of a hydrogen or helium layer (Medin \& Lai 2007), 
it is plausible that photodissociation and spallation process will
liberate ions at a sufficient rate to supply the spindown current 
(Section \ref{s:seed}).  We therefore assume a continuous charge flow
from the magnetar surface, for either sign of $\rho_{\rm GJ}$.  

The voltage that develops in this case has generally been though to 
be regulated by the gradient in $\rho_{\rm GJ}$ 
away from the neutron star surface.  The dominant contributions to
this gradient come from general relativistic
frame dragging (Muslimov \& Tsygan 1992) and magnetic field line curvature
(Scharlemann et al. 1978).   
The second effect may well be dominant in magnetars, given the strong
evidence for a non-dipolar field structure in the light curves of 
SGR flares (Thompson \& Duncan 2001).  

The main assumption here is that the 
current that is drawn from the neutron star surface can be supplied
very nearly by a flow of outgoing and returning charges,
\be
J \simeq \rho_\out v_\out + \rho_\in v_\in \simeq \left(\rho_\out -\rho_\in\right)c.
\ee
In our basic one-dimensional model,
the current density $J$ is nearly aligned with the background magnetic
field, and $\rho_\out$ and $\rho_\in$ are the densities of outgoing and
ingoing charges.  The flux of ingoing charges is provided by pair creation
within the circuit.  

We consider two basic circuit solutions.  In the first case,
the pair creation is distributed smoothly 
throughout the circuit, so as to maintain the local charge balance 
\be\label{eq:cbal}
\rho_\out(z) + \rho_\in(z) = \rhoGJ(z).
\ee
Here $\rhoGJ$ is the corotation charge density.
The second basic type of circuit solution involves
a plasma-filled gap of height $h$, which sits at the base of the circuit.  
Charges moving outward through this gap gain enough energy to emit gamma
rays that are able to convert to pairs at $z > h$, thereby creating a 
dense polarizable plasma.  The gap voltage can also be limited by
pair creation by ingoing charges, and by modes of pair creation that
are distributed throughout the volume of the gap (\S \ref{s:four}).

In a Newtonian gravitational field, the corotation charge 
density would be given by (Goldreich \& Julian 1969)
\be\label{eq:rhocorel}
\rho_{\rm GJ}^0 = -{1\over 4\pi c}{\bnabla}  \cdot ({\mathbf v}\times
{\mathbf B}) = -{\Om\cdot{\bf B}\over 2\pi c}.
\ee
This result is modified by relativistic effects:  in particular,
the rotation of the neutron star drags the frame of a zero 
angular-momentum observer with an angular velocity $\bomega = \kappa\Om$,
where the coefficient is given by\footnote{Throughout
this paper we use the shorthand $X = X_n\times 10^n$, where the value
of quantity $X$ is measured in c.g.s. units.}
\be\label{eq:kapdef0}
   \kappa(r)=\kapNS\left(\frac{r}{\RNS}\right)^{-3},  \qquad
   \kapNS=\frac{2G\INS}{\RNS^3 c^2}=0.15\,I_{{\rm NS},45}R_{{\rm NS},6}^{-3}.
\ee
Here $\INS$ is the moment of inertia of the neutron star, and $\RNS$ its radius.
This has the effect of reducing the corotation charge 
density,\footnote{The quantity on the left-hand side is, strictly,
the product of the lapse function $\alpha(r)$ and the proper charge density,
but this subtlety can be neglected in our present treatment because $B$
and the current density $\alpha\rhoGJ v$  both satisfy the same conservation equation.}
\be\label{eq:rhogjrelc}
  \rho_{\rm GJ} \;=\; \rho_{\rm GJ}^0\,\left[1-\kappa(r)\right].
\ee
The normalized charge density $\rhoGJ/B$ increases with radius
along a straight magnetic field line.

If the higher order multipole components
of the magnetic field are comparable in strength to the dipole near
the neutron star surface, then a stronger voltage can be produced than
purely by the frame dragging effect
(Barnard \& Arons 1982; Asseo \& Khechinashvili 2002).  
As we show in Appendix \ref{s:GR},
the leading contribution to the field line
curvature comes from the octopole component of the magnetic field,
and the quadrupole component to second order, 
when these two components are treated as a perturbation to the dipole,
\be\label{eq:drhogjdlb}
B{d(\rho_{\rm GJ}/B)\over dl} = {\rhoGJ\over R_C}\sin\chi\cos\phi_C 
= \K3 {\rhoGJ\over R_{\rm NS}}\left({r\over \RNS}\right)^{-3}.
\ee
Here $R_C$ is the curvature radius and 
the coefficient $\K3$ is expressed in terms of the surface
amplitudes $\varepsilon_{2,3}$ of the quadrupole and octopole via
\be\label{eq:k3def}
\K3 = \left(\varepsilon_3 + \varepsilon_2{\partial\varepsilon_2\over
   \partial\theta}\right)\,\sin\chi\,\cos\phi_C.
\ee
In addition, $\chi$ is the angle between the magnetic moment $\bmu$ and the rotation
axis $\Om$, and $\phi_C$ the angle between the plane of curvature of the magnetic
field lines and the $\bmu-\Om$ plane.

A subtlety associated with this model, which has not been fully 
explored, is that the value of $J$ that is demanded by the outer magnetosphere
need not closely match an outward charge flow of density $\rho_{\rm GJ}$ and
speed $c$.  When the current density is smaller in magnitude than $\rho_{\rm GJ}c$ 
and has the same sign, the open-field voltage remains modest close to the star,\footnote{
This statement does not apply to the outer magnetosphere, where part of the 
open-field bundle crosses the surface where $\Om\cdot\B$ vanishes;  in other words,
an outer gap can still be present even in the absence of an inner gap.}
and the mean flow speed of the outgoing charges is smaller than the speed of light
(Mestel et al. 1995; Shibata 1997; Beloborodov 2007).  We therefore focus
here on the case where $J$ is larger in magnitude than $\rho_{\rm GJ} c$, allowing for
the possibility that $J$ may have the opposite sign.
Such an imbalance could naturally result from a multipolar structure in the external
magnetic field of the magnetar; and from more subtle effects such as
resistive instabilities in the current carrying flux tube or
more exotic effects such as the loss of magnetic helicity from the closed
magnetosphere onto the open field lines.

\subsection{Voltage Solution I without an Inner Gap and Pair Multiplicity $\sim 1$}
\label{s:nogap}

We now describe a simple model for the inner part the open field circuit,
within which pair are created at the minimum rate needed to balance the
gradient in $\rhoGJ$ along the magnetic field.   This zone extends
from the neutron star surface out to a maximum radius $R_*$, which
is determined self-consistently by the stability of the solution
(\S \ref{s:svol}).   The range of radii considered here is much
broader than the width of the open-field bundle, 
i.e. $R_* - \RNS \gg \RNS\theta_{\rm open}(\RNS)$, 
where $\theta_{\rm open}(r) = (\Omega r/c)^{1/2}$.
The circuit can, therefore, be approximated as a thin tube.

We take the flow to be steady, with all particles moving
relativistically and charges of opposite signs moving in opposite directions.
Implicit in this setup is the assumption that one particle in
each newly created $e^\pm$ pair reverses direction nearly
instantaneously.
The flows of ingoing and outgoing charges satisfy the usual
continuity equations,
\beq
\label{eq:cont1}
  c\,\frac{d}{d r}\left(\nout S_\perp\right) =  \ndot S_\perp,  \\
  -c\,\frac{d}{d r}\left(\nin S_\perp\right) = \ndot S_\perp,
\label{eq:cont2}
\eeq
where $\ndot= (\dot{n}_{e^-}+\dot{n}_{e^+})/2$ is the rate of creation of
electrons and positrons per unit volume, and $S_\perp$ (eq. [\ref{eq:Sperp}]) 
is the cross sectional area of the open field-line bundle at radius $r$.
The gradient in the total charge density $\rho=\rho_\out + \rho_\in$
is obtained from the sum of equations~(\ref{eq:cont1}) and (\ref{eq:cont2}).
One finds
\be
\label{eq:ndot}
  c\,\frac{d}{d r}\left(\rho S_\perp\right)
       = 2e\ndot S_\perp\,{\rm sgn}(\rhoGJ),  \\
\ee
where $e$ is the magnitude of the electron charge.

As is shown below, the electric field $\Epar = {\bf E}\cdot \hat B$ 
is greatly reduced compared with the unscreened value in the absence 
of pair creation.
The small $\Epar$ is sustained by a small fractional departure of $\rho$ 
from the local corotation charge density. Therefore, 
\be\label{eq:charbal}
  \rho(r)\simeq\rho_{\rm GJ}(r).
\ee
Associated with this charge density is a net outward flow of 
particles moving at $v \simeq c$,
\be
\dot N_{\rm GJ} = {1\over e}|\rho_{\rm GJ}| S_\perp c
                \equiv {|I_{\rm GJ}|\over e},
\ee
The continuity equation (\ref{eq:ndot}) can then be written as
\be\label{eq:pairrate}
   \ndot S_\perp={1\over 2}\frac{d \dot N_{\rm GJ}}{dr},
\ee
A specific example in which pair creation
is driven by resonant scattering is described in \S \ref{s:five}.

In this toy model, the
net current along the open field-line bundle is in a steady state,
\be
\label{eq:Isum}
   \Iout+\Iin=I={\rm const},
\ee
where $\Iout=c\rhoout S_\perp$ and $\Iin=-c\rhoin S_\perp$. 
The strength of the
current is determined by conditions in the outer magnetosphere, 
and does not influence the qualitative nature of the solution.
In particular, current-driven instabilities in the outer magnetosphere
are assumed to occur on timescales much longer than the light-crossing time of
the inner circuit, so that changes in $I$ occur
adiabatically.  (This assumption will break down if the outer
magnetosphere is strongly dynamic, so that a turbulent spectrum of
modes is established by non-linear couplings between the waves:
Lyutikov \& Thompson 2005; Thompson 2008.)

Using $\rho\simeq\rhoGJ$ we also have
\be
\label{eq:Idiff}
  \Iout-\Iin=\IGJ.
\ee
Equations~(\ref{eq:Isum}) and (\ref{eq:Idiff}) imply 
\be\label{eq:Iin}
   \Iin=\frac{1}{2}\,\left(I-\IGJ\right), \qquad
    \frac{d\Iin}{dr}=-\frac{1}{2}\,\frac{d\IGJ}{dr}.
\ee
The outward current density can therefore differ significantly from $I_{\rm GJ}$
if $I$ does,
\be\label{eq:Iout}
I_\out = {1\over 2}(I+I_{\rm GJ}).
\ee

We allow for the possibility that a gap may form beyond some 
outer radius $R_*$.  This radius is determined, 
in the case of a magnetar, by the condition that resonantly scattered X-ray
be able to convert directly to electron-positron pairs, and corresponds
to the surface where $B=4\BQ$ (\S \ref{s:five}).  Beyond this gap,
a dense polarizable plasma is present and the solution continues
to hold in a time-averaged sense.  

\subsection{Voltage Solution II with an Inner Gap and Pair Multiplicity $\gg 1$}
\label{s:thingap}

In a super-QED magnetic field, 
the conversion of gamma rays to pairs generally occurs over a small distance
compared with the width of the open magnetic 
field lines.   We therefore focus on a gap with height smaller than $\sim
 \theta_{\rm open} R_{\rm NS}$.   Particles flow relativistically through
nearly the entire gap width, and so $\rho_\in$
and $\rho_\out$ are taken to be constant.  The electrostatic potential
satisfies the equation
\be\label{eq:pois}
{\partial^2\Phi\over\partial z^2} = -4\pi(\rho_\out + \rho_\in - \rho_{\rm GJ}).
\ee
Close to the surface of the neutron star, the
corotation charge density has an approximately linear dependence on height,
\be
\rhoGJ = \rhoGJ(R_{\rm NS})\left(1+A{z\over R_{\rm NS}}\right).
\ee
The dimensionless quantity $A$ can be expressed in terms of the
frame-dragging coefficient $\kappa$ (eq. [\ref{eq:kapdef0}])
and the octopole-quadrupole coefficient $\K3$ (eq. [\ref{eq:k3def}])
via 
\be\label{eq:aval}
A \simeq 3\kappa(\RNS) + \K3.
\ee
(see Appendix \ref{s:GR} for further details).
The pair creation rate is assumed to be zero at at height $z < h$ above
the magnetar surface; and to become large at $z > h$.  In other words,
the density of pairs that are created outside the gap is assumed to become
much larger than the change in $\rho_{\rm GJ}/e$ across the gap,
within a small distance $z-h \ll h$ from the upper boundary of the gap.

Integrating eq. (\ref{eq:pois}) with respect to $z$ with the boundary
conditions $\Phi = d\Phi/dz = 0$ at $z=0$ and $d\Phi/dz = 0$ at
$z = h$ implies a unique value of the charge density inside the gap,
\be
\rho = \rho_\out + \rho_\in = \rhoGJ\left(z = {1\over 2}h\right).
\ee
The gap solution is
\ba\label{eq:sol}
\rho_\out &=& {1\over 2}\left[{J\over c}+\rhoGJ\left({h\over 2}\right)\right];\nn
\rho_\in &=& {1\over 2}\left[-{J\over c}+\rhoGJ\left({h\over 2}\right)\right];\nn
E_z(z) &=& -{d\Phi\over dz} \;=\; 2\pi A\,\rhoGJ(\RNS) {h^2\over \RNS}
\left[{z\over h} - \left({z\over h}\right)^2\right];\nn
\Phi(z) &=& \Phi_{\rm gap}
\left[3\left({z\over h}\right)^2 - 2\left({z\over h}\right)^3\right];\nn
\Phi_{\rm gap} &\equiv&-{\pi\over 3} A \rhoGJ(\RNS) {h^3\over \RNS}.
\ea
These expressions for $E_z$ and $\Phi$ are very similar to the diode solutions
constructed by Arons \& Scharlemann (1979) for a space-charge limited flow
in a planar geometry, the main difference being that a returning flow
of charges is included.  

The establishment of the screened state at $z>h$ typically requires
a very modest additional potential drop $\Delta\Phi$ compared with the gap voltage
(\ref{eq:sol}).  Outgoing charges of the sign opposite to $\rhoGJ$ must
reverse direction so as to maintain $\rho_\in + \rho_\out = \rhoGJ(r)$.  
The outgoing charge density has both positive and negative components,
$\rho_\out = \rho_\out^+ + \rho_\out^-$, with $|\rho_\out^+|$, $|\rho_\out^-|
\gg |\rho_\in|$.  The density of ingoing charges is just sufficient
to compensate the difference between $J$ and $\rhoGJ c$,
\be\label{eq:rhoin}
\rho_\in(z) = -{1\over 2}\left[{J\over c} - \rhoGJ(z)\right].
\ee
When the pair creation rate is high at $z > h$, there is a small step in the
ingoing charge density at $z = h$,
\be\label{eq:drhoin}
\rho_\in(h+\epsilon) - \rho_\in(h-\epsilon) \;=\; 
{1\over 2}\left[\rhoGJ(h)-\rhoGJ\left({h\over 2}\right)\right]
\;\simeq\; {A\over 4}\left({h\over \RNS}\right)\,\rhoGJ(\RNS).
\ee

We work out a specific example of such a gap in \S \ref{s:four},
where pair creation is dominated
by the resonant scattering of thermal X-rays.  When the surface
magnetic field is $\sim 10^{14}$ G, particles crossing the gap acquire
an energy $e|\Phi_{\rm gap}| \sim 10^3-10^4\, m_ec^2$ (depending on the
surface temperature).  

\subsubsection{Sustaining the Low-Voltage Circuit Solution
Outside the Gap}

The gap solution (\ref{eq:sol}) depends on the ability of the circuit
to maintain a configuration with a relatively high particle density
outside the gap, 
\be
n_\out + n_\in \gg n_{\rm GJ} \equiv {|\rhoGJ|\over e},
\ee
and an approximately charge-balanced state with a low voltage,
\be\label{eq:rhobal}
\rho_\out(r) + \rho_\in(r) \simeq \rhoGJ(r).
\ee
In the inner part of the circuit, this second condition can be
satisfied if some fraction of the `negative' charges (with a sign
opposite to $\rhoGJ$) reverse direction.  This is possible because
the outgoing primary charges that flow through the gap will create
a dense cloud of secondary pairs as they lose energy to resonant
drag.  The energy spectrum of these secondary particles extends
far below the gap voltage.  As a result,
only a small additional voltage is needed to create the required
number of ingoing charges.  Indeed, the photons
that are resonantly scattered by the outgoing primary charges
will continue to create pairs to a much larger distance from the star,
where $B \sim 4\times 10^{12}$ G and the energy of the secondary
pairs is $\sim 10-100$ times smaller than $e\Phi_{\rm gap}$
(\S \ref{s:resdrag}).  

Screening of the electric field near the outer gap boundary is straightforward
in the solution (\ref{eq:sol}), because $E_\parallel \rightarrow 0$ inside
the upper boundary of the gap even in the absence of pair creation.  The
minimal pair density that is needed for screening is higher when $E_\parallel$
at the first point of pair creation remains comparable to the peak value 
in the circuit (e.g. Shibata, Miyazaki, \& Takahara 1998, 2002).  This is
the case in the vacuum gap model of Ruderman \& Sutherland (1975).

\begin{figure}
\epsscale{0.7}
\plotone{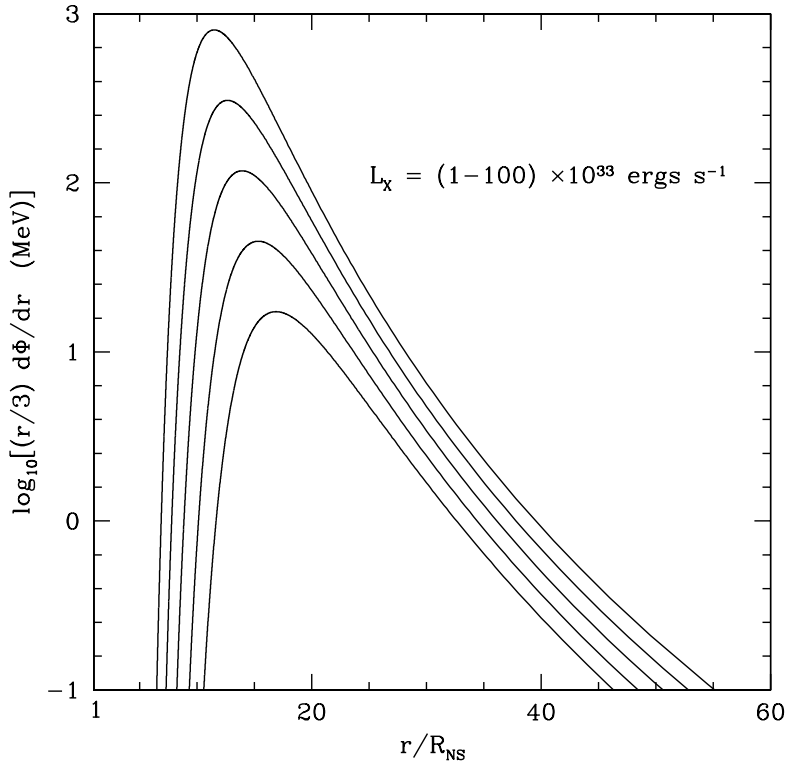}
\caption{
Electric potential gradient (given by eq. [\ref{eq:phicrit}])
that is required to cancel the radiative force exerted by thermal
X-rays on a stationary electron or positron.  A blackbody spectral
distribution is assumed, with emission from the surface of the
neutron star ($\RNS = 10$ km) in a single polarization mode.  The
different curves correspond to X-ray luminosities $(1,3,10,30,100)\times
10^{33}$ ergs s$^{-1}$ (from bottom to top).  
If a surface gap emits dense clouds of outgoing pairs, then
an electric potential close to this value will develop
beyond the radius $\sim (10-15)\RNS$ where the potential drop
$e (r/3) d\Phi/dr$ is larger than the minimum energy of the outgoing
charges.  This relatively weak
electric force allows the outflowing plasma to maintain the local
charge balance $\rho = \rhoGJ(r)$ in this zone.
\vskip .2in\null
}
\label{f:pot}
\end{figure}

At greater distances from the star, where the cyclotron energy
of an electron is $\sim \kB\Tbb$, mildly relativistic electrons
and positrons feel a very strong drag force (e.g. Thompson \&
Beloborodov 2005).  In this zone, an outgoing charge can reverse direction only
in the presence of a relatively strong $E_\parallel$.  
For the moment, we will neglect the effects of radiation transfer
and assume that a large fraction of the outflowing X-rays are unscattered.
Near the magnetic pole, an electrostatic potential 
\be\label{eq:phicrit}
e\,d\Phi = -e E_\parallel dr =
          8.0\times 10^2\,B_{\rm NS,15}^{-1/3} R_{\rm NS,6}^{-1}
 \left[{(dL_X/d\omega_X)d\omega_X\over 10^{35}~{\rm ergs~s^{-1}}}\right]\,
  \left({\hbar\omega_X\over {\rm keV}}\right)^{-2/3}\qquad {\rm MeV}
\ee
just cancels the outward force of resonant scattering on the electrons.
The dependence of $\Phi$ on radius is plotted in Fig.
\ref{f:pot}, assuming a blackbody X-ray spectrum.
When the magnitude of this potential is
much larger than the kinetic energy of the secondary pairs, it is 
not possible for the circuit to maintain a steady state in which
only a fraction of the secondary pairs reverse direction.

A second steady-state circuit solution can, however, be constructed
at a somewhat smaller voltage.  Here we drop the assumption that
all the charges move close to the speed of light.  An electric potential
just slightly smaller in magnitude than eq.
(\ref{eq:phicrit}) will allow the negative charges
to flow {\it outward} at a speed somewhat below the speed of light,
even while the outgoing
positive charges remain highly relativistic.  The proportions of the 
current $J_+$, $J_-$ that are carried by the positive and negative
charges can be left arbitrary, and related to the respective space charge
densities via $J_+ \simeq \rho_+ c$, $J_- = \rho_- v_-$.
Demanding that $J_+ + J_- = J$ and that condition (\ref{eq:rhobal})
be satisfied, we find
\be
{v_-\over c} = \left(1 + {J-\rhoGJ c\over |J_-|}\right)^{-1}
= {|J_-|\over J_+ - \rhoGJ c}.
\ee
This drift speed is indeed smaller than the speed
of light, since we are assuming that the total current density $J > \rhoGJ c$
(otherwise, the negative charges are not required to maintain local
charge balance within the circuit).  The transition between the
two circuit solutions occurs at the radius 
where $eE_\parallel r \sim \gamma_{\rm min} m_ec^2$, where
$\gamma_{\rm min}$ is the minimum Lorentz factor of the secondary
pairs.  


The fluctuations in the current density are not small when the twist on the
open magnetic field lines is dynamic, and there is a significant coupling 
between the ingoing and reflected torsional waves.   The resulting cascade
generates very strong current fluctuations, and most of the energy in
the fluctuating toroidal magnetic field is deposited in the particles
(Thompson \& Blaes 1998).
In this case, the creation of pairs does not depend on large scale
electric fields, and very high pair multiplicities are possible.  This
type of circuit solution is examined in Thompson (2008).

Finally, we note that our gap solution makes sense only if the outgoing
and ingoing charges within the gap have opposite signs.  One sees from eq.
(\ref{eq:sol}) that this is possible only if $J$ is larger in magnitude than
$\rho_{\rm GJ}(R_{\rm NS})c$.  When $J$ is opposite in sign to $\rho_{\rm GJ}
c$ but smaller in magnitude, then the current can be supplied by an inward
subluminal drift of the corotation charge, but only if there is a source of
charge in the outer magnetosphere, e.g, associated with an outer gap, or a
diversion of current across the magnetospheric boundary.

\subsection{Supply of Positive Charges from the Surface of a Magnetar}
\label{s:seed}

The current on the open magnetic field lines
 is supplied mainly by electrons when the
spin vector and dipolar magnetic field are aligned
($\Om\cdot {\bf B} > 0$ and $\rho_{\rm GJ} < 0$); 
and by positive charges
(ions or positrons) in the case of anti-alignment 
($\Om\cdot {\bf B} < 0$ and $\rho_{\rm GJ} > 0$).  
It has been suggested that the observational manifestations of
magnetars could be substantially different in the two spin orientations,
because heavy ions would be tightly bound to the surface of
the star in long molecular chains (Zhang \& Harding 2000).  Although
the binding energy of atoms heavier than oxygen is indeed too
high to allow thermionic emission (Thompson et al. 2000),
two mechanisms of creating unbound ions are available.  

First, downward-moving relativistic electrons trigger a cascade of pairs
and gamma rays below the condensed surface.  Gamma rays with energies
of $\sim 20$-30 MeV are created in such a cascade, and 
are able to knock out protons and neutrons from heavier nuclei.  
On average $\sim 0.1(\gamma/10^3)$ proton is created per 
seed relativistic electron of Lorentz factor $\gamma$
(Beloborodov \& Thompson 2007).  
The pairs created in a surface gap (where $B < 4\BQ$) have
typical Lorentz factors $\gamma \sim 10^4$ (\S \ref{s:four}), and the magnitude
of the returning charge density can approach $\rhoGJ$ if
there is a significant mis-match between $J$ and $\rhoGJ c$.
In this case, knockout appears marginally effective at supplying
a Goldreich-Julian flux of particles.  However, the energy of the returning charges
is regulated to $\gamma \sim 10^3$ if $B_{\rm NS} > 4\BQ$ (\S \ref{s:five}).  
A gap may therefore form above the condensed surface with a voltage regulated to the
value that gives about $\sim 1$ proton knocked out per returning charge.  This voltage
($\sim 10^{10}$ V) remains too small for the accelerated particles to emit
curvature photons in the gamma ray band.
An additional population of energetic charges could possibly be created in the
outer magnetosphere, with Lorentz factors high enough that
returning to the star they suffer negligible resonant drag.  (The net kinetic
luminosity of these returning charges need only reach $\sim 10^4 m_ec^2
I_{\rm GJ}/e$, typically $\sim 10^{-3}$ of the spindown luminosity
of a magnetar.)

Second, ions can be photodissociated from the molecular
chains in the presence
of a strong flux of 10-100 keV photons.  Most active magnetars are observed to
emit such hard X-rays at a rate $\dot N_X \sim 10^{42}-10^{43}$ s$^{-1}$ 
(Kuiper et al. 2004; Mereghetti et al. 2005).  
The observed spindown of the neutron star implies
a much smaller particle flux along its open field lines:  
a spindown luminosity $L_{\rm sd}$
corresponds to a total Goldreich-Julian flux
$\dot N_{\rm GJ} \sim 1\times 10^{31}\,L_{\rm sd,33}^{1/2}$
s$^{-1}$.  This flux arises from a fraction $\pi \RNS/cP = 
2\times 10^{-5}\,(P/6~{\rm s})^{-1}$ of the neutron star surface.  
If even one part in a million of the hard X-ray flux is reflected
toward a part of the stellar surface that contains the polar
cap region, then the photodissociation rate in the polar cap could
be high enough to sustain $\dot N_{\rm ion} \sim \dot N_{\rm GJ}$.
To put this number in perspective, it will be recalled that
a twist angle of $\sim \Delta\phi$ radians in the closed
field line region of a magnetar supplies an optical depth
$\sim \Delta\phi$ at the electron cyclotron
resonance. 


Ions themselves are not efficient radiators, so the question then
arises as to how pair creation can be spawned in the open field-line
region when $\rho_{\rm GJ} > 0$.  The same issue is encountered in 
electrodynamic models of radio pulsars (e.g. Cheng \& Ruderman 1977).
The simplest answer is that a state in which the `excess' corotation
charge density $\rho_{\rm GJ} - \rho_{\rm GJ}^0$ (eq. [\ref{eq:rhogjrelc}])
is supplied by pair creation is self-sustaining, just as it is in the 
twisted magnetosphere of a magnetar (Beloborodov \& Thompson 2007).
The details of how this happens are the subject of \S\S \ref{s:four}
and \ref{s:five}.

\section{Gap in a Magnetic field $B < 4\BQ$}\label{s:four}

The strength of the surface magnetic field has an important
influence on the mechanism of pair creation (Beloborodov \& Thompson
2007; Thompson 2008). Magnetars are bright thermal X-ray sources,
so that relativistic charges produce pairs in copious amounts
by the resonant scattering of the X-rays.   When $B > 4\BQ$,
some resonantly scattered photons are created above the threshold
for pair creation, and there is essentially no delay between the positions
of scattering and pair conversion.  This leads to effective
screening of the component of electric field parallel to $\B$.

The critical field strength for direct pair creation of this type
is obtained easily by considering the Landau de-excitation of
an electron (or positron) and the energy of the resultant gamma ray.
In a frame in which the excited particle is at rest, the gamma ray
energy is maximized when it is emitted orthogonal to ${\bf B}$.
In this case, the electron receives no recoil, and energy conservation
gives $(1+2B/\BQ)^{1/2}m_ec^2 = E_\gamma + m_ec^2$.  Requiring
$E_\gamma > 2m_ec^2$ corresponds to $B > 4\BQ$.

\begin{figure} 
\epsscale{1.0} 
\plotone{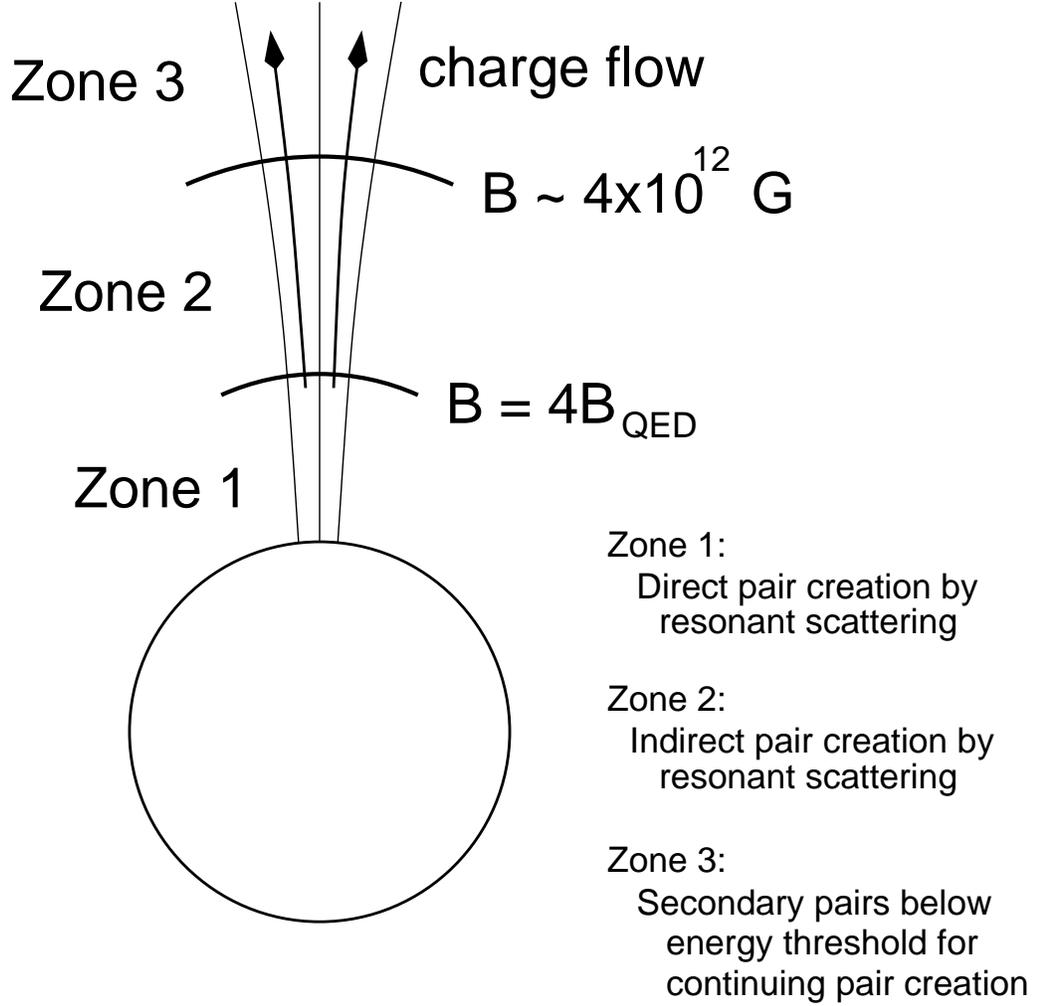} 
\caption{ 
Zones of pair creation on the magnetic field lines extending to a large 
distance from the magnetar surface.  The direct conversion of resonantly  
scattered photons to pairs is restricted to the innermost Zone I, where  
$B > 4\BQ$.  Where the magnetic field is weaker, resonantly scattered  
photons convert to pairs only after propagating a macroscopic distance. 
When a dense cloud of pairs forms in a gap 
close to the surface, large scale electric fields are screened, but new 
pairs can continue to form as the outflowing particles lose energy to  
resonant drag, out to the radius where $B \simeq 4\times 10^{12}$ G.
Secondary modes of pair creation 
are important inside the gap when the surface temperature 
$\kB T_{\rm bb} \ga 0.2$ keV.   Target photons  
that are non-resonantly scattered with a reduced cross section can  
convert directly to pairs in Zone 2, and collisions between resonantly 
scattered photons and X-rays are important at somewhat higher surface 
temperatures.  See the text for further details. 
\vskip .2in\null
} 
\label{f:zones} 
\end{figure}

In weaker magnetic fields,
direct pair creation by non-resonant scattering is possible
at a much reduced cross section.  The target photon must have
an energy larger than the resonant energy $\hbar\omega_{\rm res}'
= (B/\BQ)m_ec^2$ in the rest frame of the scattering charge.
We consider the effects of this process in \S \ref{s:quench}.
Collisions between gamma rays and thermal X-rays can also be
important, and are examined in \S \ref{s:gxcolb}.
The resulting zonal structure is summarized in Fig. \ref{f:zones}.
Many more details and references about pair creation 
in ultrastrong magnetic fields are given in Thompson (2008).

In this section we consider the case where $B < 4\BQ$ at the surface
of the star, so that the conversion of {\it resonantly} scattered
gamma rays to pairs occurs only adiabatically.
First, the gamma ray propagates in
the curving magnetic field and converts to bound positronium;
and, second, the positronium atom is dissociated when one of the constituent
charges resonantly scatters an X-ray or absorbs an infrared
photon.

\subsection{Resonant Scattering of Thermal X-rays}

The mean free path for the creation of gamma-rays, and then
for the dissociation of the positronium atom,
is obtained from the rate of resonant scattering.  An X-ray of frequency 
\be\label{eq:res}
\omega_X' = \gamma(1-\beta\mu)\omega_X = {eB\over m_ec}
\ee
in the rest frame of an electron
or positron is scattered with an enhanced cross section.  
Here $\mu$ is the direction cosine of the 
target photon with respect to the magnetic field (defined so that
$\mu > 0$ and $\beta > 0$ when the photon and charge move away from
the star).  The target photon is strongly
aberrated into a direction nearly parallel to ${\bf B}$ when
the charge moves relativistically ($\beta = v/c \simeq 1$), so that
\be\label{eq:sigres}
\sigma^{\rm res}(\omega_X') \simeq {2\pi^2\over m_ec} \delta\left(\omega_X'-
{eB\over m_ec}\right)
\ee
(Daugherty \& Ventura 1978).  We focus on the case where
the target photons are emitted spherically from the
surface of the star, with a temperature $\Tbb$ and a single
polarization state (e.g. Silantve \& Iakovlev 1980).
The effects of light bending are neglected, so that $\mu$ uniformly fills
 the interval $1 > \mu > \mumin$, where
\be\label{eq:mumin}
\mumin  = (1-\RNS^2/r^2)^{1/2}.
\ee
One has $\mumin = 0$ at the surface of the star, and 
$\mumin \simeq \RNS^2/2r^2$ at $r \gg \RNS$.  

A charge moving radially outward with a Lorentz factor
$\gamout \gg 1$ scatters X-rays at the rate (Sturner 1995)
\be\label{eq:Gresout}
\Gresout = 
{\alf\Th^3\over 2(B/\BQ)}(1-\mumin)^2\yout^2\left|\ln(1-e^{-\yout})\right|
{c\over\lbar}.
\ee
Here 
\be\label{eq:resout}
\yout \equiv {\hbar\omega_X\over \kB\Tbb} = {B/\BQ\over \gamout\Th(1-\mumin)},
\ee
is the dimensionless frequency of the target photon at $\mu = \mumin$.
We sometimes use the more compact notation
\be\label{eq:resoutb}
\yout \equiv {\gamout^\Theta\over (1-\mumin)\gamout} = 
2.3\times 10^4\,(1-\mumin)^{-1}\,B_{15}\,
\left({\kB\Tbb\over 0.5~{\rm keV}}\right)^{-1}.
\ee
Let us evaluate expression (\ref{eq:Gresout}) at the surface of the star.
An outoing charge experiences a large number of scatterings if
the target photons are drawn from the black body peak ($\yout \sim 3$),
\be\label{eq:Gresout2}
{\RNS\over c}\Gresout(\RNS) \;=\; 3.9\times 10^3\,B_{NS,15}^{-1}
\left({\kB\Tbb\over 0.5~{\rm keV}}\right)^3\,
\yout^2\left|\ln(1-e^{-\yout})\right|.
\ee

An analogous expression is obtained for the scattering rate of
an ingoing charge,
\be
\label{eq:Gresin1}
  \Gresin=\frac{\alf\Th}{2\gamin^2}\,\left({B\over\BQ}\right)\,
   \ln\left[\frac{1-\exp\left(-\frac{B/\BQ}{\gamin(1+\mumin)\Th}\right)}
                  {1-\exp\left(-\frac{B/\BQ}{2\gamin\Th}\right)}\right]
    \,\frac{c}{\lbar},
\ee
which reduces to
\be
\label{eq:Gresin2}
\Gresin={\alf\Th^3\over B/\BQ}\,\left(1-\mumin\right)\,f(\yin)\,
   \frac{c}{\lbar}
\ee
far from the stellar surface. Here $f(\yin) = \yin^3(e^{\yin}-1)^{-1}$,
\be
\label{eq:youtb}
   \yin(\gamin)={\gamin^\Theta\over\gamin},
\ee
and
\be\label{eq:gaminth}
\gamin^\Theta = {B/\BQ\over 2\Th} = 1.2\times 10^4\,B_{15}\,
\left({\kB\Tbb\over 0.5~{\rm keV}}\right)^{-1}.
\ee

\subsection{Basic Gap Solution}

We now determine the basic properties of the gap solution outlined in \S 
\ref{s:thingap}.  In this solution, the total current density
is allowed to exceed $\rhoGJ c$, with a returning flux of charges of the
sign opposite to $\rhoGJ$ supplying a significant fraction of $J$. 
The gap voltage is then proportional to the gradient in $\rhoGJ$ along the open
magnetic field lines,
\be\label{eq:Phig}
|\Phi_{\rm gap}| = {\pi\over 3}{d\rhoGJ\over dr}h^3  \equiv 
A{\pi\over 3}{\rhoGJ h^3\over \RNS},
\ee
just as it is in models of pulsar circuits with a steady, space-charge
limited flow (Arons \& Scharlemann 1979).   In contrast with models
of pulsar polar caps with realistic pair creation (e.g.
Hibschman \& Arons 2001a),
the thickness $h$ of the gap can self-consistently be assumed small compared 
with the width of the open magnetic field lines, $h \ll \theta_{\rm open}
\RNS$.   

\begin{figure}
\epsscale{1.0}
\plotone{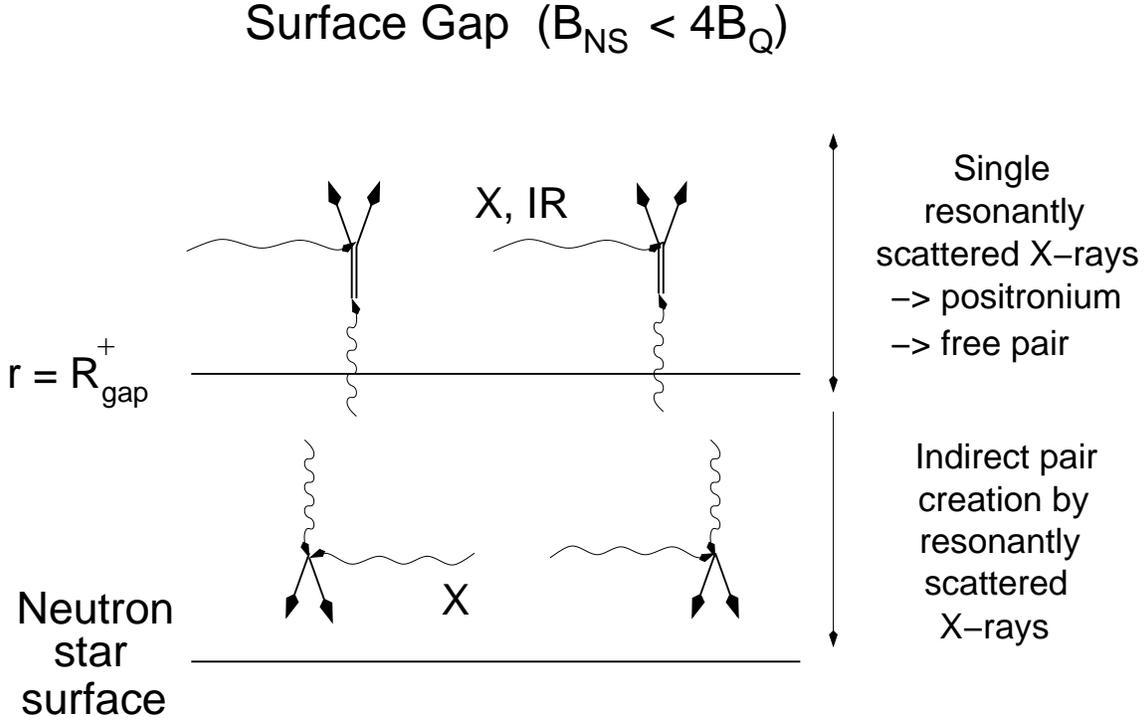}
\caption{
Structure of a gap at the surface of a neutron
star with a polar magnetic field $\BNS < 4\,\BQ$.
Some key modes of pair creation are labelled.
Outgoing particles resonantly upscatter thermal X-rays
that are emitted from the surface of the neutron star.
When the surface temperature is lower than $\sim 0.2$ keV,
the width of the gap is is set by the conversion of
these gamma rays to free pairs off the magnetic field.
(The gamma ray first converts to bound positronium, which is
then rapidly dissociated when one of the charges resonantly
scatters an ambient X-ray.)   A fraction of the
gamma rays within the gap can convert to pairs by
colliding with an X-ray before reaching the threshold
energy for single-photon pair creation, and direct pair
creation by non-resonant scattering is also important.
The gap voltage is lowered by these effects when 
$\kB\Tbb \ga 0.2$ keV.
\vskip .2in\null
}
\label{f:gap}
\end{figure}

We start with the simplest solution, in which $\Phi_{\rm gap}$ and $h$
are determined by i) the gradient in $\rhoGJ$ away from the neutron
star surface; and ii) the degree of curvature of the magnetic field lines
(Fig. \ref{f:gap}).
This solution is valid if pair creation occurs mainly by the adiabatic
conversion of gamma rays off the magnetic field, and the charges
streaming through the gap feel a negligible drag force due to resonant
scattering.  It turns out that both of these conditions are satisfied
if the surface temperature is below a critical value, $\kB\Tbb \la
0.15-0.2$ keV.

\begin{figure}
\epsscale{1.0}
\plotone{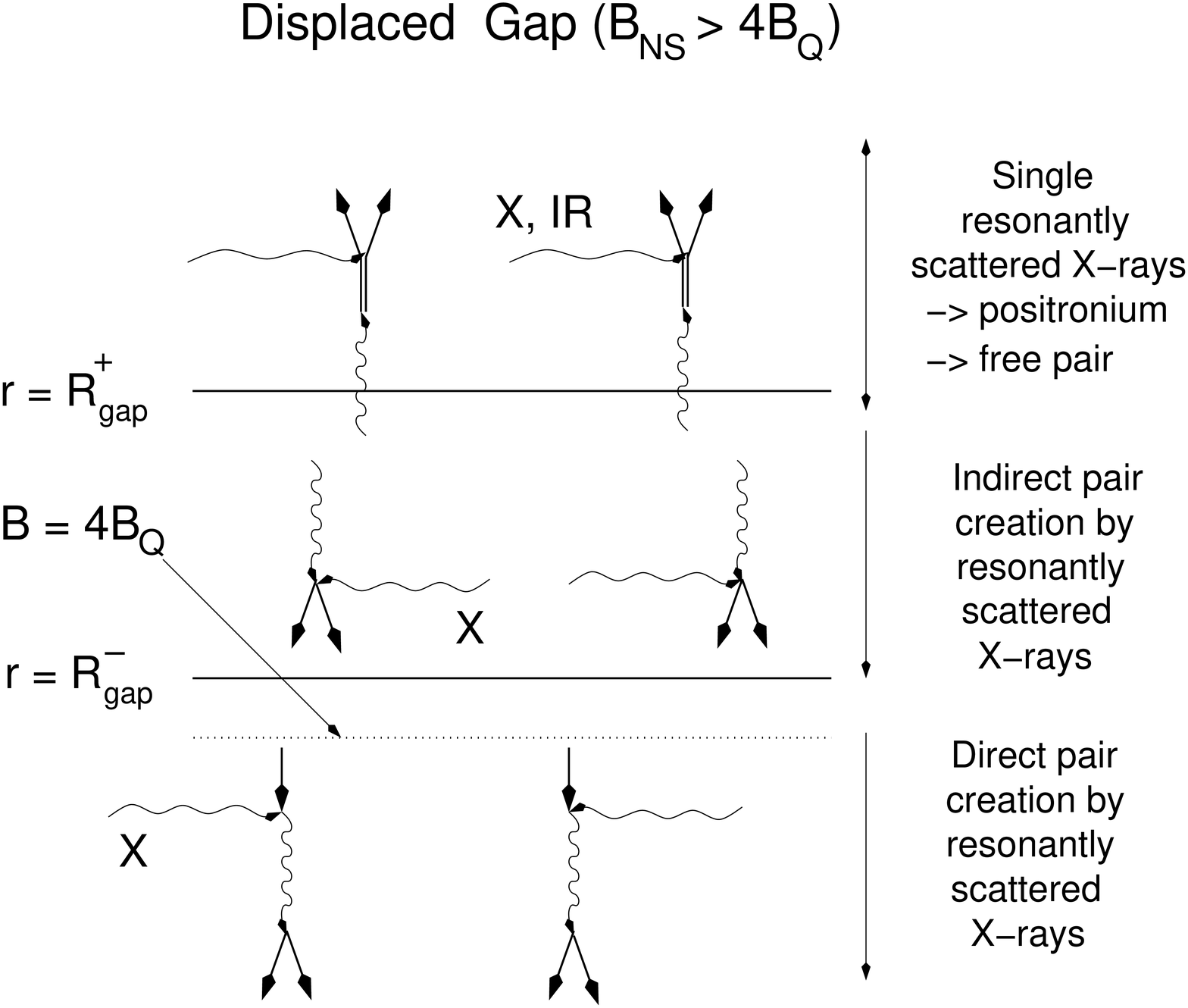}
\caption{
Possible structure of a gap whose inner boundary is
shifted just outside the surface $\BNS = 4\,\BQ$,
in the case of a magnetar with a polar magnetic
field $\BNS > 4\,\BQ$.  The existence of such a gap
depends on the non-linear development of the two-stream
instability within the circuit.
\vskip .2in\null
}
\label{f:gap2}
\end{figure}

The thickness of the gap is determined by summing three quantities,
and then minimizing this sum:  
i) the distance $\Delta z$ for an outgoing particle to be accelerated
to the Lorentz factor $\gamma$ at which it resonantly scatters a background
photons;  ii) the distance $\Delta z'$ for the resonantly scattered photon
to convert to a pair; and iii) the distance $\Delta z''$ for one of the 
created particles to reverse direction in the background potential, thereby
polarizing the plasma within the gap.   This procedure is self-consistent
when the mean number of scattered photons per outgoing charge satisfies,
\be\label{eq:mgam}
{\cal M}_\gamma \gg {\Delta\rho_\in\over\rhoGJ} = {Ah\over 4\RNS}.
\ee
Here $\Delta\rho_\in = \rho_\in(h+\epsilon) -\rho_\in(h-\epsilon)$ is 
the shift in the ingoing charge density
across the outer boundary of the gap, and is given by eq. (\ref{eq:drhoin}).
The right-hand side of eq. (\ref{eq:mgam}) is small, and so this condition
is consistent with weak drag.   In this case, the gap thickness
depends only weakly on the density of target photons, and we can 
take the energy of the resonantly scattered photon to be the maximum
possible energy.  In a super-QED magnetic field, this is a sizeable
fraction of the kinetic energy of the scattering charge,
\be
E_\gamma({\rm max}) = {2B/\BQ\over 1+2B/\BQ}\gamma m_ec^2.
\ee

In the regime of weak drag, multiple scatterings are rare and
a scattering particle attains its kinetic energy
$\gamma m_ec^2$ at a potential drop $\Delta\Phi \simeq \gamma m_ec^2/e$
from the gap boundary.
After conversion of the gamma-ray to a pair, one of the two outgoing charges
will reverse direction after the particles pass through an additional
potential drop $\Delta\Phi'' \simeq {1\over 2}\gamma m_ec^2/e$.   The
relation between particle energy and position within the gap is
\be\label{eq:gamvz}
{\gamma\over\gamma_{\rm max}} \;=\; 3\left({z\over h}\right)^2 - 2\left({z\over h}\right)^3
\;\equiv\; \phi\left({z\over h}\right),
\ee
where
\be
\gamma_{\rm max} \;=\; 
{A\over 6}\left({\Omega\RNS\over c}\right){eB_{\rm NS}\RNS\over m_ec^2}\,
\left({h\over\RNS}\right)^3 \;\equiv\; K_{\rm gap}\,
\left({h\over \RNS}\right)^3.
\ee
(see eq. [\ref{eq:sol}]).    The position of scattering is a distance
$\Delta z$ from one boundary of the gap, and the position of pair conversion
is a distance $\Delta z''$ from the other boundary, where
\be\label{eq:dzrel}
\phi\left({\Delta z\over h}\right) = {\gamma\over\gamma_{\rm max}};
\qquad
\phi\left({\Delta z''\over h}\right) \simeq {\gamma/2\over\gamma_{\rm max}}.
\ee

As the thickness of the gap shrinks,
so does the voltage and the particle Lorentz factor.   
The delay between gamma ray emission and pair conversion therefore plays an
essential role in stabilizing the gap.  
The conversion distance $\Delta z'$ for the gamma ray
scales inversely with $\gamma$, and so increases with decreasing $h$,
\be\label{eq:dzpr}
\Delta z' \;=\; {2m_ec^2 \over E_\gamma({\rm max})}\,R_C = {2\over \gamma}
\left({2B/\BQ\over 1+2B/\BQ}\right)^{-1}R_C.
\ee
Here $R_C$ is the curvature radius of the magnetic field lines, which in
the case of strong field curvtaure is given by eq. (\ref{eq:rcnonlin}).
The equilibrium gap width can be obtained by setting
\be\label{eq:delmin}
{\partial h\over\partial\gamma} = {\partial\over\partial\gamma}\left(\Delta z + 
\Delta z' + \Delta z''\right) = 0.
\ee
Differentiating eqs. (\ref{eq:dzrel}) and (\ref{eq:dzpr}) with respect to $\gamma$,
and substituting into eq. (\ref{eq:delmin}) gives
\be
{1\over\phi'(\Delta z/h)} + {1\over 2\phi'(\Delta z''/h)}
 \;=\; {\Delta z'/h\over \phi(\Delta z/h)}.
\ee
Combining this with the relations $\phi(\Delta z/h) = 2\phi(\Delta z''/h)$
and $\Delta z' = h - \Delta z - \Delta z''$,, we obtain the solution
\be
\Delta z \simeq 0.38\,h;  \qquad \Delta z'' \simeq 0.25\,h,
\ee
and
\be
\gamma_{\rm max}
 \;=\; 17.5\,\left({2B/\BQ\over 1+2B/\BQ}\right)^{-1}\,{R_C\over h}.
\ee
Substituting again for $\gamma_{\rm max}$ gives the equilibrium gap width,
\ba\label{eq:hgap}
{h\over\RNS} &=& 
\left[{K_{\rm gap}\over 17.5}
\left({2B/\BQ\over 1+2B/\BQ}\right)\right]^{-1/4}\nn
&=& {2.7\times 10^3\over {\cal F}^{1/4}(B)}\,
{R_{\rm NS,6}^{1/2}\over A^{1/4} B_{\rm NS,14}^{1/4}}
\,\left({P\over 6~{\rm s}}\right)^{1/4}
\left({R_C\over \RNS}\right)^{1/4}\qquad{\rm cm},
\ea
and voltage
\be\label{eq:gmax}
\gamma_{\rm max} = {e|\Phi_{\rm gap}|\over m_ec^2} = 
{6.5\times 10^3\over {\cal F}^{3/4}(B)}\,A^{1/4} B_{\rm NS,14}^{1/4}
R_{\rm NS,6}^{1/2}\,\left({P\over 6~{\rm s}}\right)^{-1/4}
\left({R_C\over \RNS}\right)^{3/4},
\ee
where 
\be
{\cal F}(B) \equiv {2B/\BQ\over 1+2B/\BQ}.
\ee

We now consider the number of photons that are resonantly upscattered
by the outgoing charges.  The scattering rate is, in fact, somewhat
higher for the ingoing charges (which see photons moving toward them
nearly head on).  The maximum energy of the resonantly scattered
photons is, however, the same for particles moving in both directions,
and the distribution of kinetic energies within the gap is symmetric
between the two directions.\footnote{Each outflowing particle 
scatters X-rays at a modest rate.  We therefore neglect the
effect of the recoil on the motion of a typical outgoing particle.}
The gap width, as determined by the
minimization procedure (\ref{eq:delmin}), is therefore independent
of the direction in which the scattering charges are moving.  

A simple estimate of the energy of the gamma ray
is obtained by viewing the scattering process as
resonant absorption followed by de-excitation.  The absorbing charge
decelerates from a Lorentz factor $\gamma$ to
$\gamma_1 < \gamma$, which is found from the equation of
energy conservation, $\hbar\omega_X+\gamma m_ec^2=\gamma_1 E_B$,
where $E_B = (1+2B/\BQ)^{1/2}m_ec^2$ is the energy of the first Landau
level.  This gives (neglecting the small energy $\hbar\omega_X$)
\be
 \label{eq:gam22}
  \gamma_1 = \frac{\gamma m_ec^2}{E_B}
  = \gamma\,\left(1+2{B\over\BQ}\right)^{-1/2}.
\ee
The final Lorentz factor of the electron, when it returns to the
ground state, equals $\gamma_1$ only if the de-excitation photon 
is emitted perpendicular to $\bB$ in the electron frame. Otherwise
there is a recoil effect whose sign depends on the photon direction.
If we adopt $\gamma_1$ as an estimate of the mean final Lorentz factor,
the typical loss in scattering is\footnote{
It equals $\frac{2}{3}\,\gamma$ for $B = 4\,\BQ$ and 
$\simeq 0.85\,\gamma$ for $B = 10^{15}$~G. The lost energy is carried 
by the scattered photon. 
It is shared between two new particles when the photon converts to a pair.
Each particle in the created pair therefore has a typical kinetic energy
$\sim 0.3$-$0.4\,\gamma\,m_ec^2$ in the direction of the scattering charge.}
\be\label{eq:frecoil}
\Delta\gamma \simeq  \gamma-\gamma_1 
\equiv f_{\rm recoil}\gamma,
\ee
where
\be
f_{\rm recoil} =  1-\left(1+2{B\over\BQ}\right)^{-1/2}.
\ee

\begin{figure}
\epsscale{1.25}
\plottwo{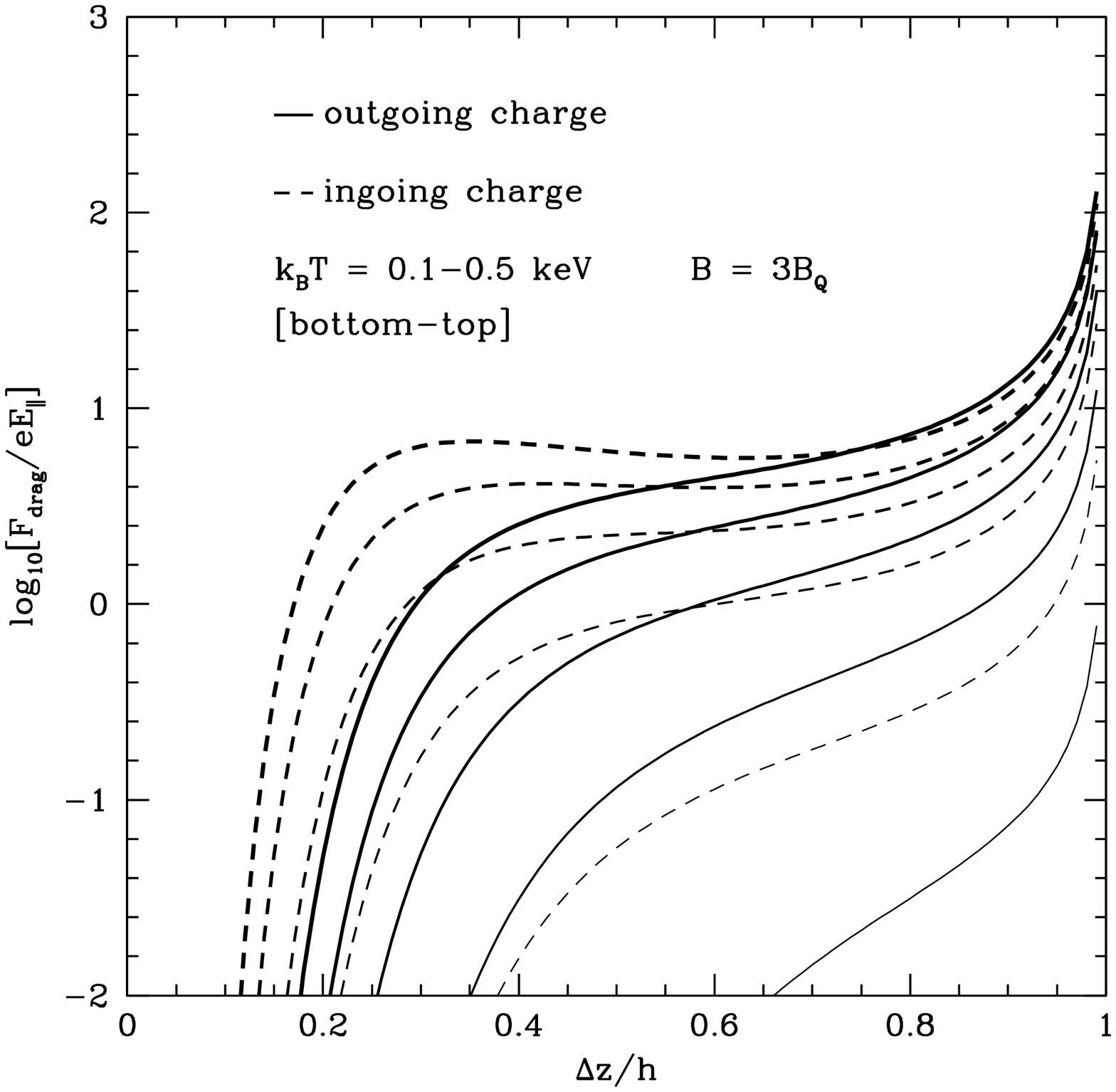}{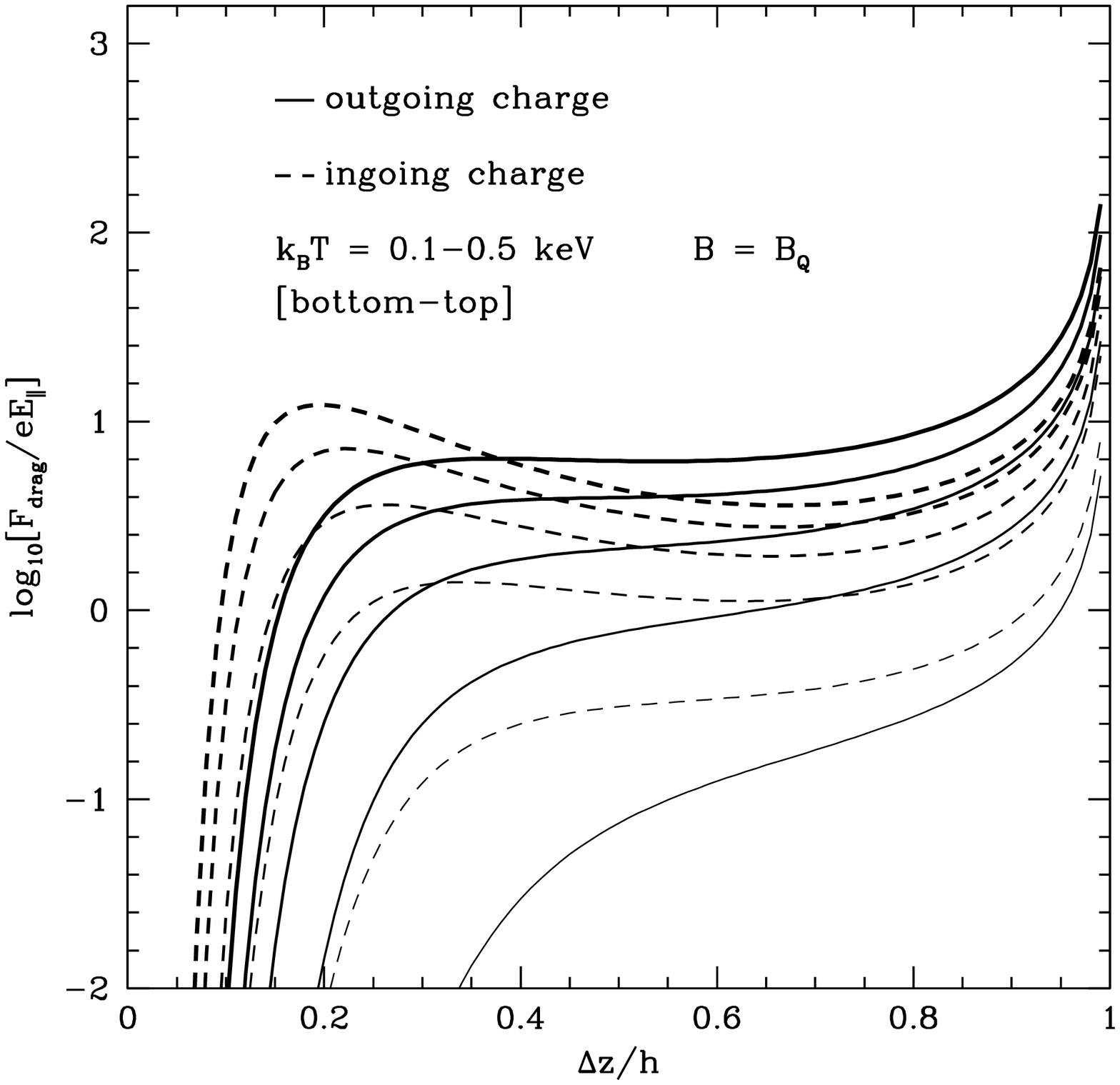}
\caption{Ratio of drag force due to resonant scattering, and
the accelerating electric force, within the gap solution given by 
eq. (\ref{eq:sol}).
The forces on the outgoing and ingoing charges are evaluated
seperately, as a function of the distance from the injection
radius ($z = 0,h$ respectively).   The results are displayed for
surface magnetic fields $\BNS = \BQ$ and $3\BQ$.  
Each set of five curves corresponds to blackbody temperatures
$k_BT_{\rm bb} = 0.1,0.2,0.3,0.4$, and 0.5 keV (thicker curves representing
higher temperatures).
We choose parameters $A = 1$, $\RNS = 10$ km, $R_C = \RNS$
and $P = 6$ s.
\vskip .2in\null
}
\label{f:drag}
\end{figure}

\begin{figure}
\epsscale{0.7}
\plotone{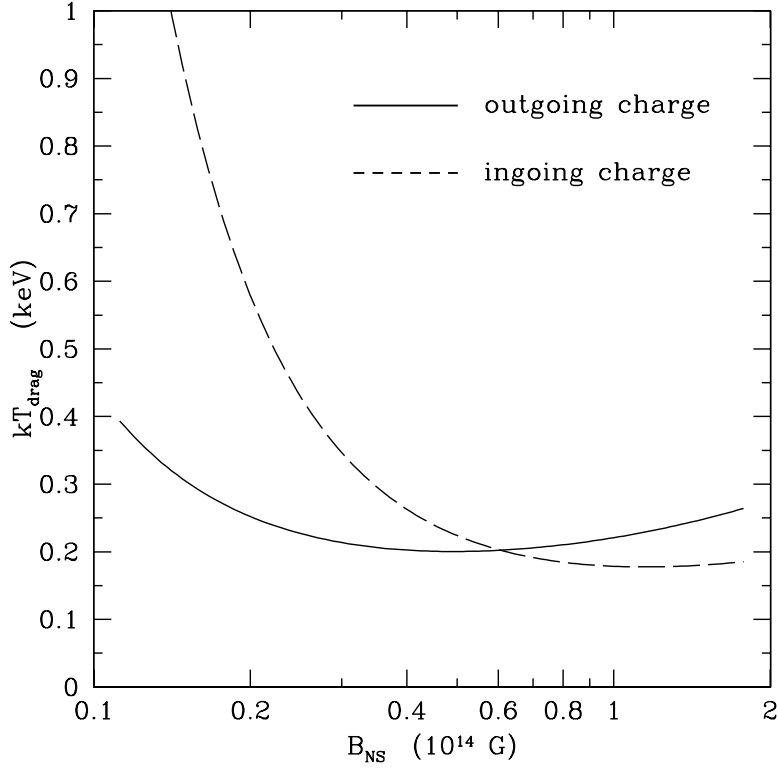}
\caption{
Critical blackbody X-ray temperature above which charges flowing
through the middle of the gap (\ref{eq:sol}) feel a drag force
equal to $eE_\parallel$, as a function of magnetic field in the gap.
We choose parameters $A = 1$, $\RNS = 10$ km, $R_C = \RNS$
and $P = 6$ s.
\vskip .2in\null
 }
\label{f:dragB}
\end{figure}

The number of scatterings per outgoing particle is distributed over gamma ray
energy $E_\gamma = f_{\rm recoil} \gamma m_ec^2$ according to
\be
E_\gamma {d{\cal M}_\gamma\over dE_\gamma}\biggr|_\out = 
\gamma {d{\cal M}_\gamma\over d\gamma}\biggr|_\out
= {\gamma m_ec\Gresout \over e E_z},
\ee
where we take $\mumin = 0$ at the surface of the star. 
Substituting eqs. (\ref{eq:sol}), (\ref{eq:Gresout}), 
(\ref{eq:hgap}), and (\ref{eq:gmax})  gives 
\ba\label{eq:nmulsc}
E_\gamma {d{\cal M}_\gamma\over dE_\gamma}\biggr|_\out
 &=& 1.46\,{\alf\Th^2\over \gamma_{\rm max}^2\,{\cal F}(B)}\,
\left({R_C\over\lbar}\right)\,
{\yout|\ln(1-e^{-\yout})|\over (z/h) - (z/h)^2}\nn
&=& {6.2\,{\cal F}^{1/2}(B)\over A^{1/2}\,B_{\rm NS,14}^{1/2}}\,\left({\kB\Tbb\over0.5~{\rm keV}}\right)^2\,
\left({P\over 6~{\rm s}}\right)^{1/2}\,\left({R_C\over \RNS}\right)^{-1/2}\,
{\yout|\ln(1-e^{-\yout})|\over (z/h) - (z/h)^2},
\ea
where the normalized energy of the target photon is
\be\label{eq:youtg}
\yout = 
     {B/\BQ\over\gamma\Th} 
     = 0.36\,\left({\gamma_{\rm max}\over\gamma}\right)\,
       {{\cal F}^{3/4}(B)\,B_{\rm NS,14}^{3/4}\over A^{1/4} R_{\rm NS,6}^{1/2}}
     \left({\kB\Tbb\over 0.5~{\rm keV}}\right)^{-1}\,
     \left({P\over 6~{\rm s}}\right)^{1/4}\,
     \left({R_C\over\RNS}\right)^{-3/4}.
\ee
The analogous expression
for ingoing charges is obtained from eq. (\ref{eq:Gresin2}) for
the scattering rate $\Gresin$;  one finds
\be\label{eq:nmulscb}
E_\gamma {d{\cal M}_\gamma\over dE_\gamma}\biggr|_\in
\;=\; \left[{\ln(1-e^{-\yout/2})\over\ln(1-e^{-\yout})}-1\right]\;
E_\gamma {d{\cal M}_\gamma\over dE_\gamma}\biggr|_\out,
\ee
where we have made use of the relations $\yin = {1\over 2}\yout$
for $\mumin = 0$ (eqs. [\ref{eq:resout}], [\ref{eq:youtb}]).

A useful measure of resonant scattering within a gap is the relative
magnitude of the electric force and the drag force due to scattering,
\be
{F_{\rm drag}^{\out,\in}\over eE_\parallel} \;=\; f_{\rm recoil}
\,{d\Mgam\over d\ln E_\gamma}\biggr|_{\out,\in}.
\ee
This quantity is plotted in Fig. \ref{f:drag} as a function of
blackbody temperature for polar magnetic fields $\BNS = \BQ$ and $3\BQ$.
The mean energy of the resonantly scattered photons is reduced by
drag when $F_{\rm drag}/e|E_\parallel| \ga 1$ at $z \sim {1\over 2}h$. 
The critical temperature $T_{\rm drag}$ above which this inequality
is satisfied is plotted in Fig. \ref{f:dragB}.  Although the drag
formally is important when $\kB\Tbb \ga 0.2$ keV, we find that
other pair creation mechanisms will reduce the gap voltage in that temperature
range and thereby reduce the effects of drag.  

The possibility of a surface gap in a magnetic field stronger than 
$4\BQ$ deserve some comment.   In this case most resonantly scattered
X-rays will quickly convert to pairs, so that $\Delta' \simeq 0$.  
The gap structure examined here then appears to be unstable to collapse to
small $h$, and it may disappear entirely.  
On the other hand, a self-consistent
gap structure may be sustained outside the surface where
$B = 4\BQ$, with properties very similar to those described here
(Fig. \ref{f:gap2}).

\subsection{Reduction of the Gap Voltage by \\ 
Secondary Modes of Pair Creation}\label{s:pairgap}

The gap voltage can be reduced by additional modes of pair creation, which
we now consider.  Gamma rays created by the resonant scattering of thermal X-rays
can collide with the X-rays to form free pairs, 
before they reach the threshold for single-photon pair creation
(see Zhang \& Qiao 1998 for a discussion of this process in the context
of ordinary radio pulsars).
Higher energy gamma rays will also be emitted even when $B < 4\BQ$, with
a cross section comparable to the Klein-Nishina value.  The emission rate is lower
than for resonant scattering, but the conversion to a pair is almost immediate.
This second process turns out to impose the stronger
limitations on the thermal X-ray flux at low temperatures, and we 
discuss it first.

The rate of pair creation integrated through the gap must not exceed a critical value,
\be\label{eq:dnpmax}
{e\over c}\int_0^h dz\, \dot n_\pm \;<\; 
{h\over 4}{d\rhoGJ\over dr} \;\equiv\; {A\over 4}\left({h\over\RNS}\right)\,\rho_{\rm GJ},
\ee
which corresponds to a net production of negative charges equal 
to the jump (\ref{eq:drhoin}) in the ingoing charge density at the outer
boundary of the gap.  We focus here on a gap situated at the surface of the neutron
star, where outgoing and ingoing charges make significant contributions to the 
integral (\ref{eq:dnpmax}).

\begin{figure}
\epsscale{0.7}
\plotone{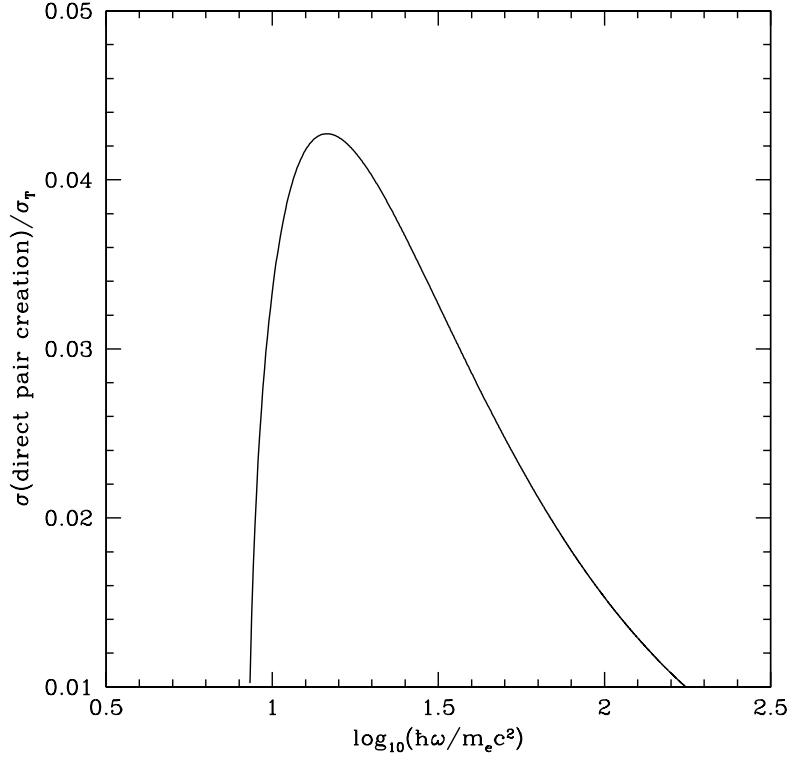}
\caption{Klein-Nishina cross section integrated over scattering
angles that result in an outgoing photon above the threshold
for direct pair creation, $\sin\theta_{kB,2}'\hbar\omega_{X,2}' > 2m_ec^2$.
The incident photon is assumed to propagate parallel to the magnetic field,
which has only a modest effect on the cross section when the photon
energy is larger than $\sim 3\hbar eB/m_ec$.   Polarization effects 
are important only when the magnetic field
is close to the threshold value $4\,\BQ$ for direct pair creation following
Landau de-excitation.  In that case, an E-mode photon has a significantly 
higher threshold energy for pair creation than does an O-mode photon.  
The threshold condition for pair creation is essentially polarization
independent when $B \ll 4\,\BQ$.
\vskip .2in\null}  
\label{f:kn}
\end{figure}

\subsubsection{Direct Pair Creation through Non-resonant Scattering}
\label{s:quench}

We now work out the kinematic constraints on direct pair creation
by non-resonant scattering.  The cross section for scattering of a
gamma ray by an electron is well approximated by the Klein-Nishina
expression when $\omega\ga 3eB/m_ec$ (Gonthier et al. 2000).  The rate of
Klein-Nishina scattering of peak photons is given by (Blumenthal \& Gould 1970)
\be\label{eq:GresoutKN}
\Gamma^{\rm KN}_\out = (1.65){\alf^2\Th^2\over 4\pi\gamma}
\ln\left({\gamma\Th}\right){c\over\lbar}
\ee
(assuming that the thermal photons are emitted in a single polarization
mode).  Relative to the rate of resonant scattering of low-energy photons
from the Rayleigh-Jeans tail, this is
\be
{\Gamma^{\rm KN}_\out\over\Gresout} \sim 2\times 10^{-3}\,\left({\gamma\over\gamout^\Theta}\right)
{\ln[(\gamma/\gamout^\Theta)(B/\BQ)])\over\ln(\gamma/\gamout^\Theta)}
\qquad (\gamma > \gamout^\Theta).
\ee
This expression holds in sub-QED magnetic fields for $\gamma > \gamout^\Theta (B/\BQ)^{-1}$.

In the rest frame of the electron, the X-ray photon has an initial energy 
$\hbar\omega_X' \simeq \gamma(1-\cos\theta_{kB})\hbar\omega_X$ and is taken 
to scatter through an angle $\theta_{kB,2}'$ with respect to ${\bf B}$. 
The threshold condition for direct pair creation following the scattering is
\be\label{eq:eperp}
\hbar\omega_{X,2}' \sin\theta_{kB,2}' > 2m_ec^2,
\ee
where 
\be
\omega_{X,2}' = {\omega_X'\over 
   1 + (\hbar\omega_X'/m_ec^2)(1-\cos\theta_{kB,2}')}
\ee
is the frequency of the photon after scattering.
The quantity on the left-hand side of eq. (\ref{eq:eperp}) is maximized when
\be
\cos\theta_{kB,2}' = {1\over 1+m_ec^2/\hbar\omega_X'}.
\ee
The minimum energy of the incident photon for direct pair creation is
\be
\hbar\omega_X' = (4+\sqrt{20})m_ec^2.
\ee
Approximating $1-\cos\thkB \simeq {1\over 2}(r/\RNS)^{-2}$
in the frame of the star gives the threshold energy
\be
\gamma = {4(2+\sqrt{5})m_ec^2\over\hbar\omega_X}\,
\left({r\over \RNS}\right)^2
\ee
for direct pair creation by a relativistic electron moving away from the star.

More energetic photons will scatter into a state that can convert
directly to an electron positron pair over some range of scattering angles,
\be
{4x\over 5} - {2\over 5}\sqrt{{5\over 4} - x^2} <
\cos\theta_{kB,2}' < {4x\over 5} + {2\over 5}\sqrt{{5\over 4} - x^2},
\ee
where $x \equiv 1 + m_ec^2/\hbar\omega_X'$.    The Klein-Nishina cross
section integrated over this range of angles is plotted as a function
of incident photon energy in Fig. \ref{f:kn}.  One sees that 
this integrated cross section peaks 
over the range of frequencies $\hbar\omega_X' = (10-20)m_ec^2$.
Denoting it by $\sigma_{\rm n-res}$, we make the approximation that
the scattering is concentrated at a given frequency $\hbar\omega_0 \simeq
16\,m_ec^2$, so that $\sigma_{\rm n-res}(\omega_X') = 
\sigma_{\rm n-res}^0 \omega'_X \delta\left(\omega'_X - \omega_0\right)$.
Here $\sigma_{\rm n-res}^0 = 0.04\,\sigma_T$.
(It will turn out that the rate of direct pair creation
off a thermal photon field, maximized with respect to $\gamma$, is
independent of $\omega_0$.)  

The rate of direct pair creation can then be written as
\be
\Gamma_{\rm n-res}^{\out,\in} = \int d\omega_X \int_{\mumin}^1 2\pi 
   d\mukB
\,{I_{\omega_X}\over \hbar\omega_X}\,(1\mp\mu)\,
\sigma_{\rm n-res}(\omega_X,\mu,\gamma).
\ee
Here $\omega_X$ and $I_{\omega_X}$ are the frequency and 
intensity of the thermal X-rays, which we assume to be emitted
from the neutron star surface in a single polarization mode
(the extraordinary mode),
\be
I_{\omega_X} = {\hbar\omega_X^3\over (2\pi)^3c^2}\,
\left[\exp\left({\hbar\omega_X\over\kB\Tbb}\right)-1\right]^{-1}.
\ee
We evaluate this expression for values of $\gamma$ where
$\omega = \omega_0/\gamma(1\mp\mu)$ sits in the Wien
tail of the thermal X-ray distribution.  
\be\label{eq:gamnresmax}
\Gamma_{\rm n-res}^\out(r,\gamma) = {k_{\rm B}T_{\rm bb}\over\hbar}\,
\left({\omega_0^2 \sigma_{\rm n-res}^0\over 4\pi^2 c^2\,\gamma^2}\right)\,
\,\exp\left[-{\hbar\omega_0/k_{\rm B}T_{\rm bb}\over 
\gamma(1-\mumin)}\right]
\ee
for an outgoing charge, and
\be\label{eq:gamnresmin}
\Gamma_{\rm n-res}^\in(r,\gamma) = {k_{\rm B}T_{\rm bb}\over\hbar}\,
\left({\omega_0^2 \sigma_{\rm n-res}^0\over 4\pi^2 c^2\,\gamma^2}\right)\,
\,\left\{\exp\left[-{\hbar\omega_0/k_{\rm B}T_{\rm bb}\over 2\gamma}\right]
- \exp\left[-{\hbar\omega_0/k_{\rm B}T_{\rm bb}\over 
                   \gamma(1+\mumin)}\right]\right\}
\ee
for an ingoing charge.  

\begin{figure}
\epsscale{0.7}
\plotone{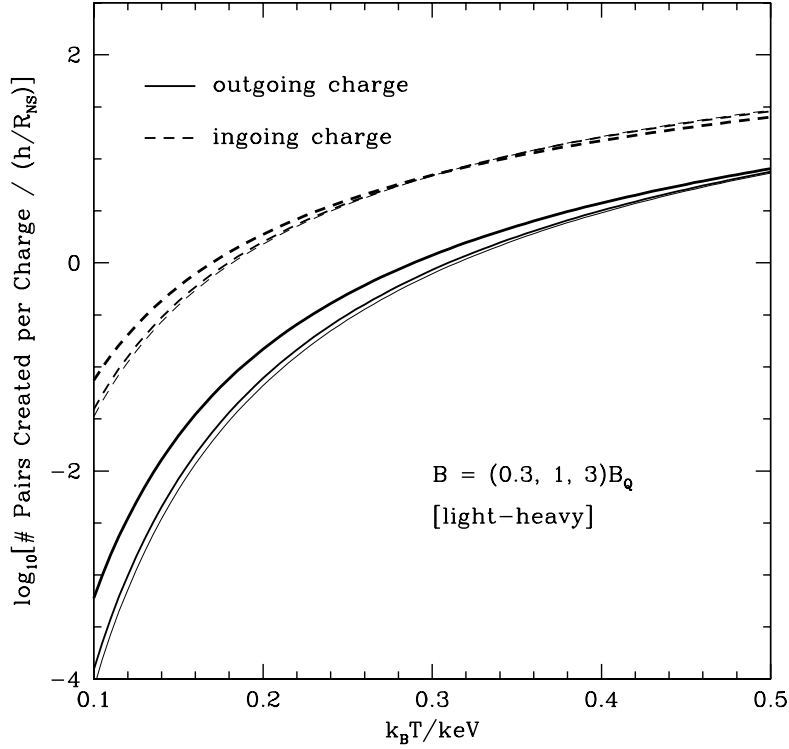}
\caption{Number of pairs that are created by the direct conversion
of non-resonantly scattered photons, within the gap given by eq. (\ref{eq:sol}).
The background magnetic field is weaker than the critical value
$4\BQ$ for direct pair creation by resonantly scattered photons.  
The contributions from outgoing and ingoing charges are evaluated
seperately.  The gap voltage is reduced by this process
when eq. (\ref{eq:screen}) is satisfied.
Each set of three curves corresponds to magnetic
fields $0.3\BQ$, $\BQ$ and $3\BQ$ (thicker curves representing
stronger fields).
We choose parameters $\sigma_{\rm n-res}^0 = 0.04\,\sigma_T$,
$\hbar\omega_0 = 16\,m_ec^2$, $A = 1$, $\RNS = 10$ km, $R_C = \RNS$
and $P = 6$ s. 
\vskip .2in\null}  
\label{f:kn2}
\end{figure}

The number of pairs that are created within the gap by each charge
is obtained by evaluating the expression
\be
N_\pm\Bigr|_{\out,\in} = \int_0^h {dz\over c}\Gamma_{\rm n-res}^{\out,\in},
\ee
where the Lorentz factor at height $z$ is given by eq. (\ref{eq:gamvz}).
The results are displayed in Fig. \ref{f:kn2}, where we have
removed a factor $h/\RNS$.

\subsubsection{Collisions between X-rays and Gamma-rays}
\label{s:gxcolb}

The rate of pair creation by photon collisions can be written in 
terms of an integral over the distributions of the gamma rays and 
the secondary target X-rays,
\be\label{eq:ggint}
\dot n_\pm\Bigr|_\out = \int d\omega_\gamma \int d\omega_X \int 2\pi d\mu_X
\left({d\Mgam\over d\omega_\gamma}\biggr|_\out\,f_\out n_{\rm GJ}\right)
\left({I_{\omega_X}\over\hbar\omega_X}\right)
\sigma_{X\gamma}(\omega_\gamma,\omega_X,\mu_X)(1-\mu_X).
\ee
Here $\omega_\gamma = E_\gamma/\hbar$ and $\omega_X$ are the frequency 
of the gamma ray and the target X-rays.  This expression applies
to the case where the gamma rays are emitted parallel to ${\bf B}$ by outgoing 
relativistic charges.  
A similar expression applies to the case where the gamma rays are
emitted by ingoing relativistic charges:  one replaces
$f_\out$ with $f_\in$ and $1-\mu_X$ with $1+\mu_X$.  Here
\be\label{eq:foutin}
f_\out \;=\; {n_\out\over n_{\rm GJ}}
\;=\; {1\over 2}\left(1+{J\over\rhoGJ c}\right);\qquad
f_\in \;=\; {n_\in\over n_{\rm GJ}} \;=\;
 {1\over 2}\left({J\over\rhoGJ c}-1\right).
\ee
The multiplicity $d\Mgam/dE_\gamma$ of resonantly scattered gamma rays
is given respectively by eqs. (\ref{eq:nmulsc}) and (\ref{eq:nmulscb}) in these
two cases.

We make the approximation that the cross section for collisions between
X-rays and curvature gamma rays is concentrated at the threshold
$\omega \omega_X (1-\mu_X) = 2m_e^2$, namely
\be
\sigma_{X\gamma}(\omega_\gamma,\omega_X,\mu_X) \;=\;
\sigma_{X\gamma}^0\; \omega_\gamma\omega_X(1-\mu_X)\;
\delta\left[\omega_\gamma\omega_X(1-\mu_X) - 
2\left({m_ec^2\over\hbar}\right)^2\right].
\ee
The value of $\sigma_{X\gamma}^0$ depends on the shape of the two
photon spectra:  following eq. (41) of Thompson (2008),
we normalize to $\sigma_{X\gamma} = 0.02\,\sigma_T$
(corresponding to a magnetic field $B = 4\,\BQ$).
The integrals over $\omega_X$, $\mu_X$ in eq. (\ref{eq:ggint}) 
are then easily completed,
giving
\be\label{eq:dnpgg}
\dot n_\pm\Bigr|_\out = f_\out n_{\rm GJ}
{\sigma_{X\gamma}^0\Th c\over \pi^2{\lbar}^3}\,
\int\, {d\Mgam\over d\omega_\gamma}\biggr|_\out\,
\left({c\over\lbar\omega_\gamma}\right)^2
\left|\ln\left[1-\exp\left(-{2m_ec^2\over\hbar\omega_\gamma\Th}\right)
\right]\right|\,d\omega_\gamma,
\ee
where we have taken $\mumin = 0$.  
The corresponding expression for the ingoing charges is
\be\label{eq:dnpggb}
\dot n_\pm\Bigr|_\in = f_\in n_{\rm GJ}
{\sigma_{X\gamma}^0\Th c\over \pi^2{\lbar}^3}\,
\int\, {d\Mgam\over d\omega_\gamma}\biggr|_\in\,
\left({c\over\lbar\omega_\gamma}\right)^2
\ln\left[{
1-\exp\left(-2m_ec^2/\hbar\omega_\gamma\Th\right)\over
1-\exp\left(-m_ec^2/\hbar\omega_\gamma\Th\right)}
\right]\,d\omega_\gamma.
\ee

\begin{figure}
\epsscale{0.7}
\plotone{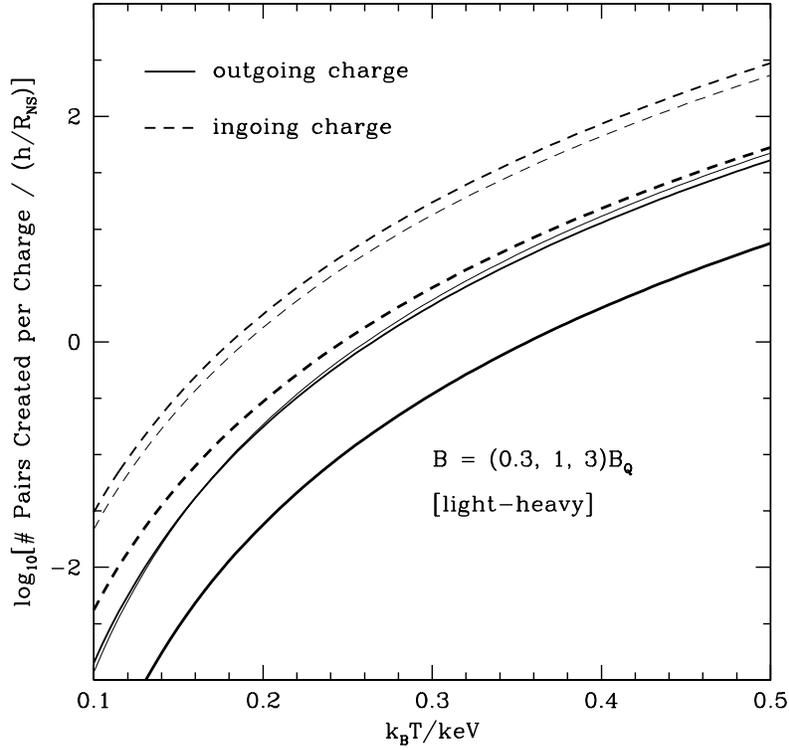}
\caption{Number of pairs created per charge flowing through a surface gap,
by collisions between thermal X-rays and resonantly scattered X-rays.
The gap solution is given by eq. (\ref{eq:sol}), with $B < 4\BQ$.
The contributions from outgoing and ingoing charges are evaluated
seperately.  The gap voltage is reduced by this process
where eq. (\ref{eq:screen}) is satisfied.
Each set of three curves corresponds to magnetic
fields $0.3\BQ$, $\BQ$ and $3\BQ$ (thicker curves representing
stronger fields).
We choose parameters $A = 1$, $\RNS = 10$ km, $R_C = \RNS$
and $P = 6$ s. 
\vskip .2in\null}  
\label{f:gamgam}
\end{figure}

To obtain the net pair creation rate per unit area of the gap,
the multiplicity $d\Mgam/dE_\gamma$
of gamma rays created by resonant scattering must be weighted
by the distance $\Delta z = h-z$ ($z$) between the position $z$ of 
resonant scattering and the outer (inner) boundary of the gap.  The 
number of pairs that are spawned within the gap by each outgoing
or ingoing charge is, then,
\be
N_\pm\Bigr|_{\out,\in} = {1\over f_{\out,\in} n_{\rm GJ}c}\,\int d\omega_\gamma
(h-\Delta z){d \dot n_\pm|_{\out,\in}\over d\omega_\gamma}.
\ee
Here $\Delta z$ is the distance from the injection
boundary.  The relation between
frequency and position within the gap is obtained by assuming
that $\hbar\omega_\gamma = f_{\rm recoil} \gamma(\Delta z) m_ec^2$, where
$\gamma(\Delta z)$ is given by eq. (\ref{eq:gamvz}), and the recoil factor
$f_{\rm recoil}$ by eq. (\ref{eq:frecoil}).   The energy $\hbar\omega_{X,0}$
of the initial target X-ray is given by
\ba\label{eq:ying}
\yout &=& {\hbar\omega_{X,0}\over\kB\Tbb} = {B/\BQ\over\gamma\Th}\nn
   &=& {f_{\rm recoil}(B/\BQ)\over\Th}\,
          \left({\hbar\omega_\gamma \over m_ec^2}\right)^{-1}.
\ea
This expression applies to an outgoing charge; similarly one has
\be\label{eq:yinh}
\yin = {1\over 2}\yout = 
{f_{\rm recoil}(B/\BQ)\over 2\Th}\,
          \left({\hbar\omega_\gamma \over m_ec^2}\right)^{-1}.
\ee
for an ingoing charge.  The integrals over $\omega_\gamma$ in eq. 
(\ref{eq:dnpgg}) and (\ref{eq:dnpggb}) can then 
be transformed to integrals over $\yout$ and $\yin$.  One finds,
\be\label{eq:pairmult}
N_\pm\Bigr|_{\out,\in} \;=\; K_{X\gamma}\,{h\over \RNS}\,I_{\out,\in}
\ee
where
\ba\label{eq:kdef}
K_{X\gamma} &\equiv& {0.15\,\alf\Th^5\over\gamma_{\rm max}^2{\cal F}(B)}
\left(f_{\rm recoil}{B\over\BQ}\right)^{-2}\,
\left({\sigma_{X\gamma}^0\RNS R_C\over {\lbar}^4}\right)\nn
&=&{26 {\cal F}^{1/2}\over A^{1/2}f_{\rm recoil}^2}\,
{R_{\rm NS,6}\over B_{\rm NS,14}^{5/2}}\,
\left({\sigma_{X\gamma}^0\over 0.02\,\sigma_T}\right)\,
\left({\kB\Tbb\over0.5~{\rm keV}}\right)^5\,
\left({P\over 6~{\rm s}}\right)^{1/2}\,\left({R_C\over\RNS}\right)^{-1/2},
\ea
and
\ba\label{eq:integral}
I_\out &\equiv& \int_{\widetilde\omega_{\out,\rm min}}^\infty d\yout {\yout^2\over (\Delta z/h)}
\ln(1-e^{-\yout})\,\ln\left[1-e^{-2\yout/f_{\rm recoil}(B/\BQ)}\right];\nn
I_\in &\equiv& 8\int_{\widetilde\omega_{\in,\rm min}}^\infty d\yin {\yin^2\over (\Delta z/h)}
\ln\left[{1-e^{-2\yin}\over 1-e^{-\yin}}\right]\,
\ln\left[{1-e^{-4\yin/f_{\rm recoil}(B/\BQ)}\over
       1-e^{-2\yin/f_{\rm recoil}(B/\BQ)}}\right].\nn
\ea
The minimum values of $\yout$ and $\yin$ are found by substituting 
$\gamma = \gamma_{\rm max}$ in eqs. (\ref{eq:ying}), (\ref{eq:yinh}).  
The numbers of pairs that are spawned within the gap by outgoing and 
ingoing charges are plotted separately in Fig. \ref{f:gamgam}.
Note that we have extracted a factor of $h/\RNS$ in the definition
of $K_{X\gamma}$.

\subsubsection{Reduction in the Gap Voltage}\label{s:gapred}

We now evaluate the effect of these secondary modes of pair creation on the gap structure.
The voltage drops below the value given by eqs. (\ref{eq:Phig}) and (\ref{eq:gmax})
when the pair creation rate inside the gap is high enough that the inequality
(\ref{eq:dnpmax}) is violated.  Self-consistent values of the gap thickness and
voltage can be obtained by summing the rates of pair creation
due to nonresonant scattering (\S \ref{s:quench}) and collisions
of gamma-rays with thermal X-rays (\S \ref{s:gxcolb}), and then
arbitrarily adjusting $\Phi_{\rm gap}$ by a factor $\varepsilon_{\rm gap}$.  
(The effect of resonant drag on the particle dynamics is neglected in calculating
the pair creation rate.)   A simple
iteration gives a unique value of $\varepsilon_{\rm gap}$ at which 
the screening condition (\ref{eq:dnpmax}) is marginally satisfied.  
The Lorentz factor at a fixed fractional depth $z/h$ in the gap
scales as $\gamma \propto \varepsilon_{\rm gap}$, and the 
thickness as $h \propto \varepsilon_{\rm gap}^{1/3}$ (eq. [\ref{eq:sol}]).  The coefficient
$K_{X\gamma}$ as defined in eq. (\ref{eq:kdef}) scales as 
$K_{X\gamma} \propto \varepsilon_{\rm gap}^{-2}$.

\begin{figure}
\epsscale{0.7}
\plotone{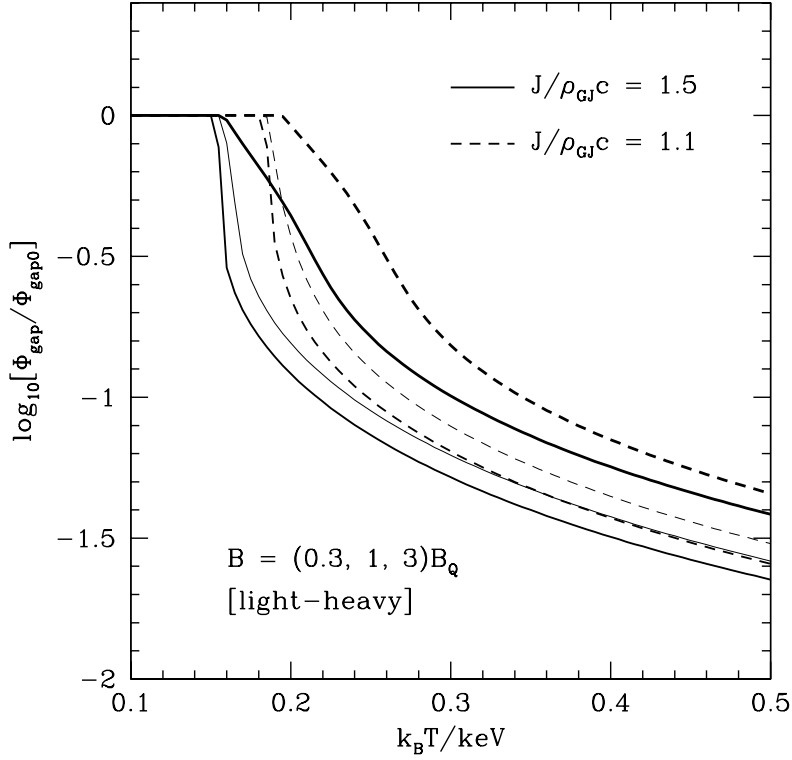}
\caption{
Reduction in the surface gap voltage $\Phi_{\rm gap}$ due to pair creation
inside the gap by non-resonant scattering of
high energy X-rays, and collisions between gamma rays
and thermal X-rays.  The unscreened gap voltage $\Phi_{\rm gap 0}$
is given by eqs. (\ref{eq:Phig}) and (\ref{eq:gmax}).
The result is given for various magnetic field strengths
and for two values of $J/\rhoGJ c$.
\vskip .2in\null}  
\label{f:voltage}
\end{figure}

\begin{figure}
\epsscale{0.7}
\plotone{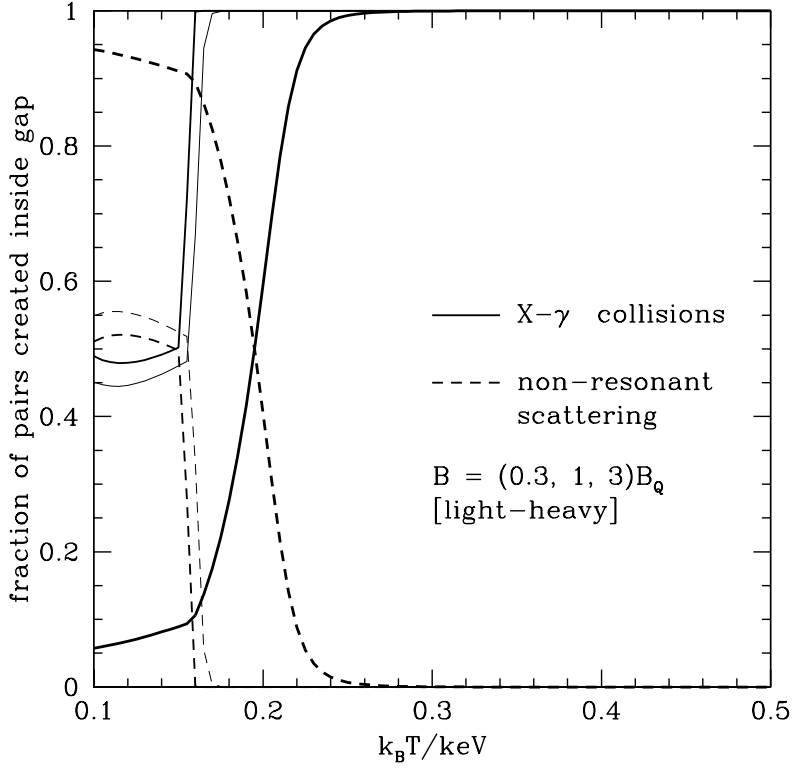}
\caption{
Proportions of the pair creation inside a surface gap due to the direct
conversion of non-resonantly scattered X-rays off the magnetic field
(\S \ref{s:quench}),  and to  collisions between gamma rays and
thermal X-rays  (\S \ref{s:gxcolb}).  The first process is
more important at temperatures below $0.2-0.3$ keV, and the second at higher
temperatures.  The current flowing through the gap is taken to be 
$J = 1.5\,\rhoGJ c$.  The break in the curves at $\kB\Tbb \sim 0.16$ keV occurs
where partial screening first becomes effective.
\vskip .2in\null}  
\label{f:voltage2}
\end{figure}


When the gap is partially screened, its voltage depends on the
the outgoing and ingoing particle densities 
$f_{\out,\in}\times n_{\rm GJ}$  (eq. [\ref{eq:foutin}]).
The screening condition (\ref{eq:dnpmax}) is readily expressed
in terms of the quantities $N_\pm\bigr|_{\out,\in}$ and $f_{\out,\in}$,
\be\label{eq:screen}
{f_\out N_\pm|_\out + f_\in N_\pm|_\in\over h/\RNS} \la {A\over 4}.
\ee
At the surface of the star,
pair creation is mainly due to the ingoing charges when $f_\in = O(1)$. 
One expects that the gap voltage is only gradually reduced
as the reverse inequality becomes satisfied:   as the energy of the target X-rays
is pushed further out onto the Wien tail, the number of targets drops
exponentially.  

This expectation is borne out in Fig. \ref{f:voltage}.
There is a critical surface blackbody temperature of about 0.15-0.2 keV
above which the gap voltage is reduced.  In the case where $\BNS = 3\BQ$,
there is a  reduction of a factor 3 at $\kB\Tbb = .23$ keV when
$J = 1.5\rhoGJ c$, and at $\kB\Tbb = 0.26$ keV when $J = 1.1\rhoGJ c$. 
A comparison of Figs. \ref{f:kn2} and \ref{f:gamgam} shows that photon collisions
are somewhat less efficient at making pairs than non-resonant scattering when
$\BNS$ is stronger than $10^{14}$ G, but are a much more important source
of pairs when $\BNS$ is close to $\BQ$ and $\kB\Tbb$ is larger than $\sim 0.2$ keV.  
The relative rates of pair creation by these two processes are plotted 
for a few values of $\BNS$ in Fig. \ref{f:voltage2}.


It should be noted that the total kinetic energy that is gained by the outgoing charges
in crossing the gap, as given by eq. (\ref{eq:gmax}) with the correction of
Fig. \ref{f:voltage}, is larger than the power radiated in gamma rays
during their transit through the gap.  As
a result, most of the pair creation that is induced by the gap occurs
{\it outside the gap}, where the primary charges are decelerated by resonant drag
(\S \ref{s:resdrag}).

\section{Cyclotron Drag on the Secondary Pairs}\label{s:cycdrag}

\subsection{Increase in Pair Multiplicity due to Continued Resonant 
Scattering}\label{s:resdrag}

Pairs that are created close to the star will continue to scatter off
ambient X-rays as they flow outward.  The resonant drag force is
stronger in more rapidly rotating neutron stars (e.g. young magnetars
or young pulsars with high X-ray luminosities), and especially in magnetars
with active, scattering magnetospheres.   The flux of pairs can
increase significantly through the conversion of the scattered X-rays,
if the initial pairs are energetic enough.  We neglect the possibility
that the pairs are continuously reaccelerated, as they would be 
if the magnetic field were strongly turbulent (Lyutikov \& Thompson 2005;
Thompson 2008).  The pair cascade then feeds off the initial kinetic 
energy of the outflowing charges, and additional generations of
pairs can be created only within a limited distance from the star.

Detailed expressions for the drag
force experienced by relativistic magnetized electrons in a radiation
field can be found in Dermer (1990), who was interested in the application
to gamma ray bursts.  Kardashev et al. (1984),  Sturner (1995),
and Lyubarskii \& Petrova (1998) consider the interaction of outgoing particles
with thermal photons near the polar cap of a neutron star, but do not
consider the interaction further out in the magnetosphere, where the
photon field is increasingly anisotropic.  

We suppose that outgoing electrons and positrons are created with $\gamout \ga
\gamout^\Theta$ (eq. [\ref{eq:resoutb}]) near the surface of the star.
A charge will lose a significant fraction of its kinetic
energy to resonant scattering over a distance $\sim r$ if 
\be\label{eq:dragcon}
f_{\rm recoil}\,\Gresout {r\over c} \sim 1,
\ee
where $f_{\rm recoil} \simeq B/\BQ$ in magnetic fields weaker than
$\BQ$ (eq. [\ref{eq:frecoil}]).  Evaluating expression
for the scattering rate off a blackbody photon
gas on the Wien tail of the distribution, one finds
\be\label{eq:yone}
{e^{\yout}\over \yout^2} = 9.8\times 10^3 {R_{\rm NS,6}\over B_{\rm NS,14}}
\,\left({B\over\BQ}\right)\,\left({\kB\Tbb\over 0.5~{\rm keV}}\right)^3
\ee
where $\yout = 
\gamout^\Theta/\gamout$.  

The resonantly scattered photons have an energy $E_\gamma \simeq \gamout
(B/\BQ)m_ec^2$.  They continue to convert to electron-positron
pairs off the magnetic field as long as
\be\label{eq:threshc}
E_\gamma \theta_{kB} \simeq E_\gamma {r\over R_C(r)} > 2m_ec^2.
\ee
The pair creation zone sits close enough to the star that we assume
that the field curvature radius $R_C$ is dominated by the quadrupole/octopole
component of the magnetic field (eq. [\ref{eq:rcnonlin}]).  Then
the threshold condition (\ref{eq:threshc}) can be expressed as
\be\label{eq:ytwo}
\yout < {\eps3\over\Th}\left({B\over\BQ}\right)^2.
\ee

We can now determine the limiting radius $R_\pm$ for pair creation, and
the characteristic energy of the outflowing particles at that radius.
The value of $\yout$ deduced from eq. (\ref{eq:yone}) varies from
$\yout \simeq 10$ near the surface of the star to $\yout \simeq 7$
at the limiting radius for pair creation. The energy of the outgoing
charges, as limited by drag, is 
\be
\gamout({\rm drag}) = {B/\BQ\over \yout(1-\mu_{\rm min})\Tbb},
\ee
and decreases from $\gamout \simeq 300(\BNS/3\BQ)(\kB T_{\rm bb}/
0.5~{\rm keV})^{-1}$ at the surface of the star.  
The magnetic field at $r = R_\pm$ works out to 
\be
B_{\rm min}(R_\pm) = 4.0\times 10^{12}\,\eps3^{-1/2}\,
\left({\kB\Tbb\over 0.5~{\rm keV}}\right)^{1/2}\qquad {\rm G}.
\ee
Hence $f_{\rm recoil} \simeq B_{\rm min}/\BQ \sim 0.1$, which means that
the last generation of pairs that is created by the conversion of 
resonantly scattered photons is well separated in energy from the 
scattering charges.  The characteristic energy of the latter has dropped to
\ba\label{eq:gam2p}
\gamout({\rm drag}) &=& {\gamout^\Theta\over\yout} = {B_{\rm min}/\BQ\over 
\yout(1-\mumin)\Th}\nn &=& 2\times 10^2\,\eps3^{-1/6}\,B_{\rm NS,14}^{2/3}\,
\left({\kB\Tbb\over 0.5~{\rm keV}}\right)^{-5/6} \qquad (r = R_\pm),
\ea
and the secondary pairs have an energy
\ba
\gamout[2] &=& {E_\gamma\over 2m_ec^2} = 
\left({B_{\rm min}\over 2\BQ}\right)\gamout({\rm drag})\nn
&=& 10\,\eps3^{-2/3}\,B_{\rm NS,14}^{2/3}\,
\left({\kB\Tbb\over 0.5~{\rm keV}}\right)^{-1/3} \qquad (r = R_\pm).
\ea

The primary outflowing charges are injected with an
energy $\gamma_0 \sim 10^3-10^4$ in the gap model described above.  
In this case, the outflowing charges multiply rapidly
just outside the gap.  Their energy spectrum can be derived
as follows (again assuming negligible reacceleration).  The mean
number of photons scattered in a radial interval $dr$ is 
$dN_\gamma = \Gresout dr/c$, and the loss of energy due to scattering
is $d\gamma = (f_{\rm recoil}\gamma)\Gresout dr/c$.  Focusing
on the zone where $B \la \BQ$, and the gamma ray carries off a
small fraction of the kinetic energy of the scattering charge,
the energy spectrum of the gamma rays and created pairs can be written as
\be
\gamout[2] {dN_\pm\over d\gamma[2]} = 
E_\gamma {dN_\gamma\over d E_\gamma} = 
\left\{ {d[f_{\rm recoil}\gamout({\rm drag})]\over 
d\gamout({\rm drag})} \right\}^{-1}.
\ee
Since $\gamout({\rm drag})$ scales approximately as $\sim r^{-1}$,
and $B(r) \propto r^{-3}$, this gives the asymptotic formula
\be
\gamout[2] {dN\over d\gamma[2]} \simeq {4 \over B/\BQ} \qquad
(B\ll\BQ).
\ee
Most of the pairs are created near
the limiting radius $R_\pm$, where the multiplicity
of pairs per primary outgoing charge has increased to
\be
{\Mpm\over \Mpm|_{\RNS}} \;\simeq\; {\gamma_0\over 2\gamout[2]}
\;=\; 500\,\eps3^{2/3}B_{\rm NS,14}^{-2/3}
\left({\kB\Tbb\over 0.5~{\rm keV}}\right)^{1/3}
\left({\gamma_0\over 10^4}\right).
\ee

A beam instability will, in fact, be triggered in this outflowing plasma,
since the distribution of particle kinetic energy extends to relatively
low values.  
The effect on the particle energy spectrum will be strongest in the outer
part of the pair creation zone, where $f_{\rm recoil} \ll 1$. 
A calculation of the energy spectrum including the combined effects
of resonant drag and the beam instability has
never been carried out, but would be very useful.

\subsection{Limiting angle for Relativistic Pair Flows}\label{s:thetmax}

In this section, we make a brief detour to address the effects of 
resonant drag in two cases:  i) the neutron star has a 
a relatively high X-ray luminosity ($L_X \ga 10^{34}$ ergs s$^{-1}$)
and a faster spin than the known magnetars ($P < 0.3$ s); and ii)
a significant fraction of the thermal X-rays are backscattered from
the magnetosphere by current-carrying charges.

\begin{figure}
\epsscale{0.7}
\plotone{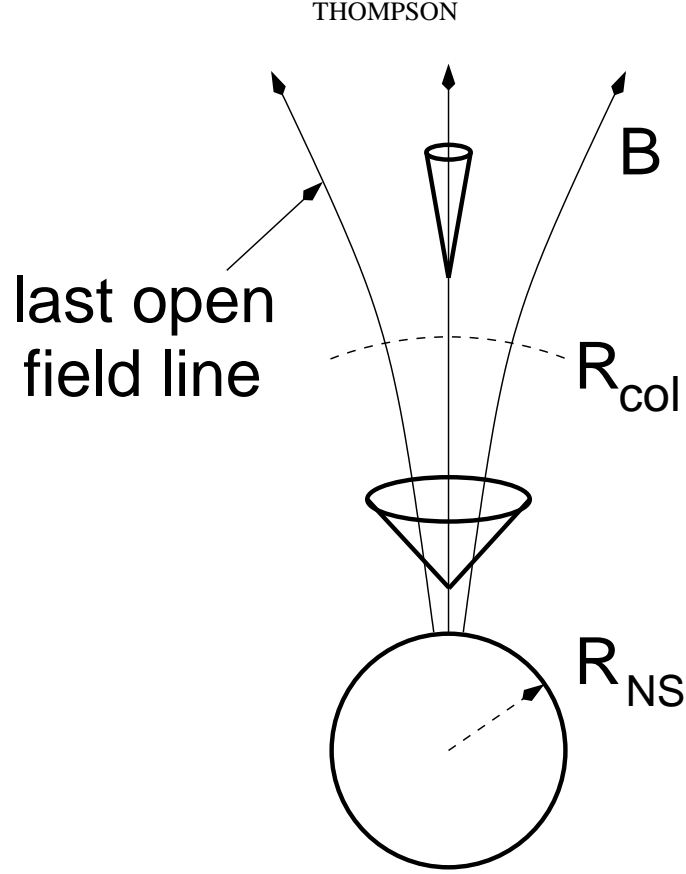}
\caption{
Close to the neutron star, the thermal X-rays
emitted from its surface fill a wide cone.  
Beyond a distance $R_{\rm col}$ (eq. [\ref{eq:rgammax}]),
the angle between the propagation direction of a photon
and the open magnetic field lines is dominated by the 
field-line curvature.
\vskip .2in\null}
\label{f:curve}
\end{figure}

The strength and sign of the drag force acting on the outgoing charges depends
on the angle $\theta_{kB}$ between their motion along ${\bf B}$,
and the target photons.  
The thermal radiation emitted from the surface of the neutron star
becomes increasingly collimated at greater distances; 
but at the same time the open dipolar\footnote{We consider the
effect of higher multipoles only when evaluating the threshold energy for
single-photon pair creation at $r<R_{\rm col}$.} 
field lines become more strongly curved.   The maximum angle value of $\theta_{kB}$
at the end of the open-field bundle is given by
\be
1-\mumin = {1\over 2}{\rm max}\left({\RNS^2\over r^2},
{\Omega r\over 4c}\right).
\ee
The first term is the large-radius expansion of eq. (\ref{eq:mumin}).
The collimation of the target photons about the local direction of the
magnetic field is tightest at a radius (Fig. \ref{f:curve})
\be\label{eq:rgammax}
R_{\rm col} = \left({4c\over \Omega \RNS}\right)^{1/3}
      \RNS
 = 12.4\,R_{\rm NS,6}^{-1/3}\left({P\over 0.1~{\rm s}}\right)^{1/3}\,\RNS.
\ee

Expression (\ref{eq:Gresout}) for the scattering rate
$\Gresout$ of the outgoing charges
is modified at $r>R_{\rm col}$.  One replaces
\be
(1-\mumin)^2 \;\rightarrow\; \left({R_{\rm NS}^2\over 2r^2}\right)\,
\left({\Omega r\over 8c}\right).
\ee
The first factor represents the continuing dilution of the target radiation
field as it flows away from the surface of the star, and the second
represents the factor $1-\mu$ in the scattering rate.  
The fraction of the particle momentum lost to drag is given by
\be\label{eq:dslow}
f_{\rm recoil}\,\Gresout {r\over c} 
= {\alf\Th^3\over 32}\left({\Omega\RNS\over c}\right)
\left({\RNS\over\lbar}\right)
\yout^2e^{-\yout}\qquad (\theta = \theta_{\rm open}).
\ee
Note that the coefficient is independent of radius.
If the spin of the star is fast enough, then the right-hand side of 
eq. (\ref{eq:dslow}) can exceed unity -- not only at $r \sim R_{\rm col}$,
but out to a very large distance from the star.  The right-hand side is
maximized at $\yout = 2$, and is greater than unity if
\be\label{eq:pdrag}
P < 0.4\,R_{\rm NS,6}^{1/2}
\left({L_X^\infty\over 10^{35}~{\rm ergs~s^{-1}}}\right)^{3/4}.
\ee
This defines the critical spin period below which the motion of the 
secondary pairs is strongly reduced by resonant scattering at $\theta
\sim \theta_{\rm open}$.  

On a closed field line that is anchored at a polar angle $\theta(\RNS) > 
(\Omega\RNS/c)^{1/2}$, expression (\ref{eq:dslow}) generalizes to 
\be\label{eq:dslowb}
f_{\rm recoil}\,\Gresout {r\over c} 
= {\alf\Th^3\over 32}\,\theta^2(\RNS)\left({\RNS\over\lbar}\right)
\yout^2e^{-\yout}.
\ee
Charges moving through the 
closed magnetosphere will experience strong drag if the guiding magnetic 
field lines are anchored at an angle larger than
\be
\theta_{\rm max}(\RNS) = 1.36\,R_{\rm NS,6}^{1/2}\,
\left({L_X^\infty\over 10^{35}~{\rm ergs~s^{-1}}}\right)^{-3/8}\qquad{\rm deg}.
\ee


\subsubsection{Young Magnetars and Neutron Stars with  \\
Fast Spins and Large X-ray Luminosities}

Let us now consider the limiting Lorentz factor of the outgoing secondary
pairs when cyclotron drag is strong at $r > R_{\rm col}$.  We focus here
on the open magnetic field lines, and therefore on stars which are spinning
faster than the known magnetars, with periods smaller than (\ref{eq:pdrag}).
The force on a scattering charge vanishes when its Lorentz factor
$\gamma = 1/\sin\theta_{kB}$.  The equilibrium value is
\be\label{eq:gammaeq}
\gamma_{\rm eq} \simeq 2\left({c\over\Omega r}\right)^{1/2} =
12.4\,\left({P\over 0.1~{\rm s}}\right)^{1/3}\,
\left({r\over R_{\rm col}}\right)^{-1/2}
\qquad (r > R_{\rm col})
\ee
when the cyclotron scattering rate is very high.
We can compare eq. (\ref{eq:gammaeq}) with the Lorentz factor which the scattering rate of
an outgoing charge is maximum,
\ba
\gamout^p &=& {\gamout^\Theta\over 2} = {B/\BQ\over 2(1-\mumin)\Th}\nn
&=& 2.2\times 10^2\; B_{\rm NS,14}R_{\rm NS,6}^{1/3}\,
\left({P\over 0.1~{\rm s}}\right)^{-1/3}\,
\left({L_X^\infty\over 10^{35}~{\rm ergs~s^{-1}}}\right)^{-1/4}
\,\left({r\over R_{\rm col}}\right)^{-4}\qquad (r > R_{\rm col}).\nn
\ea
One has $\gamout^p > \gamma_{\rm eq}$ at $r=R_{\rm col}$, so that resonant
scattering decelerates an outgoing charge at this radius.  The equilibrium
Lorentz factor is attained a bit further from the star, at the radius
\be
{R_{\rm eq}\over R_{\rm col}} = 2.3\,B_{\rm NS,14}^{2/7}R_{\rm NS,6}^{4/21}\,
\left({P\over 0.1~{\rm s}}\right)^{-4/21}\,
\left({L_X^\infty\over 10^{35}~{\rm ergs~s^{-1}}}\right)^{-1/14}
\ee
where 
\be
\gamma_{\rm eq} = \gamout^p = 8.3 \,B_{\rm NS,14}^{-1/7}R_{\rm NS,6}^{-3/7}\,
\left({P\over 0.1~{\rm s}}\right)^{3/7}\,
\left({L_X^\infty\over 10^{35}~{\rm ergs~s^{-1}}}\right)^{1/28}.
\ee
A charge is able to maintain $\gamma \simeq \gamma_{\rm eq}$ only over a modest
range in radius.  It continues to scatter off the Rayleigh-Jeans tail
of the spectrum, and then decouples from the radiation field where
the resonant frequency is $\yout^\infty \la 1$, and the asymptotic Lorentz factor is
\be
\gamma_\infty = \left({\yout^\infty\over 2}\right)^{1/7}\,\gamma_{\rm eq}(R_{\rm eq})
\sim (0.7-0.9)\,\gamma_{\rm eq}(R_{\rm eq}).
\ee

We reach an interesting conclusion:  the outflowing charges are decelerated
to a small Lorentz factor $\gamma_\infty$ at a distance $20-30\,\RNS$ from the star when
the spin period is shorter than the value (\ref{eq:pdrag}). Unless
the secondary charges are reaccelerated farther out in the magnetosphere,
the radio emission will experience a high optical depth to induced scattering;
and there would be a bias against detecting pulsed radio emission
from neutron stars with high X-ray luminosities and short spin periods.

\subsubsection{Effects of Photons Scattered in the Closed 
Magnetosphere of a Magnetar}\label{s:back}

The scattering of thermal X-ray photons by trans-relativistic charges
in the closed magnetosphere creates a second component of the 
radiation field, which is weaker but much more isotropic
(Thompson et al. 2002).  Photons from near the thermal peak
are scattered at a radius
\be\label{eq:rresb}
R_{\rm res} \simeq \left({eB_{\rm NS}\over m_ec \omega_X}\right)^{1/3}
 = 23\,B_{\rm NS,15}^{1/3}\,\left({\hbar\omega_X\over {\rm keV}}\right)^{-1/3}
\;\RNS.
\ee
This scattered radiation field can exert a strong drag force on 
outgoing charges.  We derive an upper bound on its intensity, if the
outflowing secondary charges are to avoid being decelerated to
trans-relativistic speeds.

The scattered X-rays subtend a wide range of angles, which
means that they can resonate with the outflowing charges at a relatively
small radius.  Most of the drag force is supplied by photons from 
the black body peak, and so we approximate the spectral intensity 
of the scattered radiation as
\be\label{eq:iscatt}
4\pi I_\omega \simeq 
\left({\tau_{\rm res}\over 2}\right)\,{L_X\over 4\pi 
R_{\rm res}^2(\omega_{\rm bb})}\,\delta(\omega-\omega_{\rm bb}),
\ee
where $\omega_{\rm bb} \sim 1$ keV in the case of an X-ray bright magnetar.

The drag force in the rest frame of a charge moving parallel
to ${\bf B}$ is
\be
F_\parallel' = \int d\omega'\; 2\pi\int_{\mumin'}^1
{\mu' I_{\omega'}\over c}\sigma_{\rm res}(\omega')\,d\mu',
\ee
where the scattering cross section at the cyclotron resonance
is given by\footnote{Here we assume that the charge
is moving relativistically, so that the photons are nearly collimated
with ${\bf B}$ in its rest frame.} eq. (\ref{eq:sigres}).  Transforming
to the frame of the star gives $\mu' I_{\omega'}\,d\mu'  = 
\gamma(\mu-\beta_e)I_\omega\,d\mu$, and
\be\label{eq:fpar}
F_\parallel = F_\parallel' = \gamma\,
{4\pi^3e^2\over m_ec^2}\int_{\mumin}^1 d\mu 
(\mu-\beta_e)I_\omega.
\ee
Substituting expression (\ref{eq:iscatt})
into this integral gives the 
drag time $t_{\rm drag} = \gamma m_ec/|F_\parallel|$,
\be\label{eq:tdrag}
{c t_{\rm drag}(r)\over r}
= \left({\gamma^2\over \tau_{\rm res}}\right)\,
    {64 \omega_{\rm bb}^3 R_{\rm res}^2(\omega_{\rm bb})\,m_e^2 c^2\over
  3\sigma_T B^2 L_X  r}
\ee

The scattered photons which resonate with the newly created pairs
propagate in the direction
\be\label{eq:dir}
1-\mu = {eB\over \gamma m_ec\omega_{\rm bb}},
\ee
and there is a maximum magnetic field for which a resonant
interaction is possible (corresponding to $\mu \simeq 0$),
\be
B_{\rm max} = 8.6\times 10^{12}\,\left({\gamma\over 10^2}\right)
      \left({\hbar\omega_{\rm bb}\over {\rm keV}}\right)\qquad {\rm G}.
\ee
The drag force (\ref{eq:tdrag}) at $B < B_{\rm max}$ is
\be
{c t_{\rm drag}(r)\over r}
= 0.012 \tau_{\rm res}^{-1}\,
   {B_{\rm NS,15}^{1/3}R_{\rm NS,6}\over L_{X,35}}\,
   \left({\gamma\over 10^2}\right)^{1/3}
   \left({\hbar\omega_{\rm bb}\over {\rm keV}}\right)^{2/3}
   \left({B\over B_{\rm max}}\right)^{-5/3},
\ee
and depends weakly on $\gamma$.  The resonating photons comprise 
a fraction $\sim {1\over 2}(1-\mu)$ of the backscattered flux.
The secondary pairs slow down unless $\tau_{\rm res} L_X
\la 10^{33}$ ergs s$^{-1}$.

\section{Low-Voltage Circuit with Pairs Supplied by Resonant Scattering}
\label{s:five}

We now turn to the case where the polar magnetic field of the neutron
star is stronger than $4\BQ = 1.8\times 10^{14}$ G. Close to the surface,
a significant fraction of resonantly scattered photons are above
threshold for conversion to an electron-positron pair.  The
distance $\Delta z'$ for the conversion of a gamma ray to a pair is
therefore very small.  The net effect on the surface gap structure
is to take $R_C \rightarrow 0$
in eq. (\ref{eq:dzpr}), so that its thickness (\ref{eq:hgap})
and voltage (\ref{eq:gmax}) are forced to zero.

We therefore consider the alternative circuit model, with a lower voltage,
that was outlined in \S \ref{s:nogap}.  In this model, pair creation is
smoothly distributed throughout the circuit, in such a way as to
compensate the gradient in corotation charge density.  The resonant
scattering of X-rays from the exponential tail of
the blackbody distribution can easily provide enough pairs:  if the
charges flowing through the circuit were energetic enough to 
scatter thermal X-rays from the black body peak, then the mean free path
to scattering would be be very small, $c/\Gamma^{\rm res} r 
\sim 10^{-2}-10^{-3}$ from eq. (\ref{eq:Gresout2}).

The pair creation rate per unit volume is
\be
\dot n_\pm \simeq n_\out \Gresout + n_\in \Gresin
\ee
in the zone where $B > 4\BQ$.
As before, $n_\out$ and $n_\in$ are the number densities of ingoing
and outgoing charges, and $\Gresout$, $\Gresin$ are the respective
rates of resonant scattering as given by eqs. (\ref{eq:Gresout}),
(\ref{eq:Gresin1}).  Substituting for $\dot n_\pm$ from eq. 
(\ref{eq:pairrate}), we find
\be\label{eq:scatrate}
\Gresout\,{n_\out\over n_{\rm GJ}} +  \Gresin\,{n_\in\over n_{\rm GJ}}
\;\simeq\; {c\over 2} {d\ln \dot N_{\rm GJ}\over dr}.
\ee
We further have $\Gresout \propto e^{-\yout}/\gamout^2$ and $\Gresin
\propto e^{-\yin}/\gamin^2$.  The characteristic dimensionless frequencies 
$\yout$ and $\yin$ of the target photons are given
by eqs. (\ref{eq:resout}) and (\ref{eq:youtb}), and are related by
\be
\yout = {1-\mumin\over 2}\yin.
\ee
In other words,
an outgoing charge must attain a higher energy to scatter a photon of 
a given frequency than does an ingoing charge.  This difference in
resonant energy grows with radius.  For a target photon of
energy $\hbar\omega_X = \kB\Tbb$, it is
\be
\gamout^\Theta = \frac{B/\BQ}{\Th\,(1-\mumin)}
\simeq 1.6\times 10^4\,B_{\rm NS,15}^{2/3}\,\left({B\over\BQ}\right)^{1/3}\,
\left({\kB\Tbb\over 0.5~{\rm keV}}\right)^{-1},
\ee
versus
\be
\gamin^\Theta =  \frac{B/\BQ}{2\Th}
\simeq 5\times 10^2\,\left({B\over \BQ}\right)\,
\left({\kB\Tbb\over 0.5~{\rm keV}}\right)^{-1},
\ee
at $r \gg \RNS$.

Only a fraction of the outgoing and ingoing charges must
scatter in order to satisfy eq. (\ref{eq:scatrate}), which means
that resonant drag has a small effect on the
particle energies in this circuit model. 
 The outgoing and ingoing charges pass through
comparable potential drops and therefore
have comparable energies, $\gamout \sim \gamin$.  We deduce
that $\Gresin \gg \Gresout$ because $\yout > 2\yin$ and the
scattering occurs on the exponential tail of the Planckian distribution.
Equation (\ref{eq:scatrate}) then simplifies to
\be
  2\,\Iin\Gresin=c\,\frac{d\IGJ}{dr}.
\ee
The right-hand side of this equation can be written as
$c \IGJ \,d\ln(\rhoGJ/B)/dr = A\IGJ c/r$, where the 
coefficient $A$ (eq. [\ref{eq:aval}])
includes the effects of relativistic frame dragging and magnetic 
field line curvature. The current carried by the ingoing charges
in this circuit solution is given by eq. (\ref{eq:Iin}).
The number of photons scattered per logarithm of radius works out to
\be\label{eq:screen3}
{\Gresin r\over c} =  {d\IGJ/d\ln r\over I - \IGJ}
= A\left({I\over\IGJ}-1\right)^{-1}.
\ee

One of the charges in each created pair
will typically  be decelerated by the electric field and reverse direction 
before it can scatter another X-ray.
Our toy model for the circuit assumes that this reversal is instantaneous, 
i.e. it occurs within a small distance of the point of creation.

Substituting equation~(\ref{eq:Gresin1}) for $\Gresin$ in 
equation~(\ref{eq:screen3}), and taking into account that $\yin\gg 1$,
i.e. the resonance X-rays are in the far exponential tail of the Planck
spectrum, we get the equation for $\gamin$,
\be
\label{eq:screen4}
   \frac{1}{\gamin^2}\,\frac{\alf\Th r}{2\lbar}
   \left({B\over\BQ}\right)\,
   \exp\left(-\frac{B/\BQ}{2\gamin\,\Th}\right)
    = A\left({I\over\IGJ}-1\right)^{-1}.
\ee
For typical magnetar temperatures, $\kB\Tbb\sim 0.5$~keV
($\Th\sim 10^{-3}$), the solution of equation~(\ref{eq:screen4}) 
is well approximated by
\be\label{eq:gamcrit}
    \gamin\simeq 800\,B_{15} \Thh^{-1}.
\ee
This corresponds to $\yin\sim 15$, and the scattering occurs in the 
exponential tail of the black body distribution, as expected.
As the ingoing particles move to smaller radii, 
$\gamin$ increases in proportion to $B \propto r^{-3}$.
This acceleration is sustained by an electric field $\Epar$.
The equation of motion for ingoing particles reads
\be
   -m_ec^2\,\frac{d\gamin}{dr}=e\Epar
\ee
Then we find 
\be\label{eq:Epar}
  e\Epar\simeq \frac{3\gamin\, m_ec^2}{r}.
\ee
The voltage distribution is found by integrating 
equation~(\ref{eq:Epar}),
\be\label{eq:Phi}
  -e\Phi(r)\;=\;\int_{\RNS}^r\,e\Epar\,dr \;\simeq \;
 800\,m_ec^2\, B_{\rm NS,15} \Thh^{-1}
\left[1-\left({r\over\RNS}\right)^{-3}\right].
\ee
The net potential drop between the surface of the neutron star
and the surface $B = 4\BQ$ is of the order of $1 \,B_{\rm NS,15}$ GeV.
If the primary outgoing particles are electrons ($\rhoGJ<0$), they 
are accelerated to the Lorentz factor,
\be
  \gamout(r)\simeq -\frac{e\Phi(r)}{m_ec^2}.
\ee
The outgoing particles undergo negligible scattering, and contribute
little to the net rate of pair creation, since they
see target photons that are a factor $\simeq 2$ higher in frequency
than do the ingoing charges.  We therefore expect that this circuit 
solution should apply in the inner zone for either sign of $\rhoGJ$.

The voltage (\ref{eq:Phi}) can be compared with the gap
solution (\ref{eq:gmax}), as reduced by secondary pair creation processes
(Fig. \ref{f:voltage}).  Normalizing the gap voltage to the threshold
magnetic field $\BNS = 4\BQ$, we find
\be
{\Phi(r\gg \RNS)\over \Phi_{\rm gap}(\BNS=4\BQ)} 
\;\simeq\; 
 \left({\BNS\over 4\BQ}\right)\,\left({\eps3\over 0.3}\right)^{1/2}
 \left({P\over 6~{\rm s}}\right)^{1/4}\,\times\,
\biggl\{
\begin{array}{ll}
0.05 & (\kB\Tbb = 0.1~{\rm keV}); \\ 
0.1 & (\kB\Tbb = 0.4~{\rm keV}). \\ 
\end{array}
\ee

Seeds for the ingoing $e^\pm$ discharge are supplied from the zone where
$B < 4\BQ$, and where voltage fluctuations are likely to be
larger than they are in the inner circuit.   In fact, the
true solution in both zones may be time-dependent, with repeating 
avalanches of pair creation, as is seen in a one-dimensional model of the
closed corona (Beloborodov \& Thompson 2007). The main features
of the steady circuit model should, however, 
survive in the time-dependent case:
the ingoing particles dominate the rate of pair creation in the inner
zone, that is, the pair avalanche is directed toward the star,
and the time-averaged voltage is comparable to 1~GeV.

\subsection{Stability of Voltage Solution I}\label{s:svol}

We explained in the preceding section why the surface gap described in
\S \ref{s:four} cannot be maintained when $\BNS > 4\BQ$.  We now 
address the stability question from the opposing angle:  
when $\BNS < 4\BQ$, does a gap spontaneously form in a circuit 
that is set up in the low-voltage solution,  with smoothly distributed
pair creation?   We approach this question here by considering the
stability of electrostatic modes in the circuit, but note that other
degrees of freedom could be involved.  In particular, a 
high-frequency current fluctuation would be 
generated by a turbulent cascade if the magnetic twist received
a low frequency perturbation of a moderate amplitude 
(Lyutikov \& Thompson 2005; Thompson 2008).

The instability in question bears some resemblance to the 
formation of a double layer within a current-carrying plasma.  In an 
electron-ion plasma, such a structure is believed to be seeded by a 
two-stream instability, such as an ion-acoustic instability, which is 
driven by the drift motion of the electrons that supplies the current.
Localized wavepackets that are formed via this instability undergo 
non-linear growth, leading to the formation of plasma holes which are 
bounded by layers of positive and negative charge.  Multiple double layers
can form within a single circuit, as is observed in the Earth's magnetosphere.
A diode structure with a deeper potential drop may form
through the coalescence a large number of weak double layers.  

The low-voltage circuit consists of counterstreaming clouds of electrons
and positrons, and so we first review what is known about the non-linear 
evolution of the two stream instability in a strongly magnetized pair
plasma.  Near the surface of a neutron star, the electron cyclotron
frequency is many orders of magnitude larger than the plasma frequency,
and so filamentation instabilities are suppressed.  Guidance about
the behavior of the two-stream instability can therefore be provided
by one-dimensional simulations of the purely electrostatic instability.
The perturbation will, in practice, have an electromagnetic component 
because plasma oscillations on different field lines have different phases.

We imagine preparing the ingoing and outgoing charges with a distribution
function concentrated at momenta $p_\parallel  = \pm \bar\gamma \bar V m_e$ 
parallel to ${\bf B}$.   For simplicity, we focus on the case where
$\rhoGJ \rightarrow 0$ (e.g. the open flux tube is anchored near the
rotational equator).  Then the plasma frequencies of the electrons and
positrons are equal, and are given by
\be
\omega_P^2 = {4\pi n_{e^-}\over m_e} = {4\pi n_{e^+}\over m_e} =
{2\pi Je\over m_ec},
\ee
where the densities refer to the frame of the star.  The dispersion 
relation is obtained from the longitudinal part of the dispersion tensor,
\be
1 - {\omega_P^2\over \bar\gamma^3(\omega + k\bar V)^2} -
{\omega_P^2\over \bar\gamma^3(\omega - k\bar V)^2} = 0.
\ee
The fastest growing mode has a growth rate and wavenumber
\be\label{eq:growth}
\Gamma_* = {\omega_P\over 2\bar\gamma^{3/2}}; \;\;\;\;\;
k_* = 3^{1/2}{\Gamma_*\over\bar V},
\ee
and a vanishing real frequency.  Purely growing modes exist 
within the range of wavenumbers 
$k_* < k < 2^{3/2}\Gamma_*/\bar V = 1.6\,k_*$.   In practice,
there is an imbalance in the densities of positive and negative
charges, so that the growing modes have real frequencies
of the order of $\Gamma_*$ and finite phase speeds $V_{\rm ph} \simeq c$.  

Simulations of colliding relativistic plasma clouds in 
one dimension\footnote{These simulations are of two colliding electron-ion
clouds, in which the ions supply a positive neutralizing charge distribution
and do not strongly influence the growth of the instability.}
(e.g. Dieckmann 2005, Dieckmann et al. 2006) show the growth of electrostatic
modes with a modest range of $k$ centered on $k_*$,
which then develop into a broad spectrum
of electrostatic turbulence on a timescale $\simeq 50\, \Gamma_*^{-1}$.
A necessary condition for the existence of a beam-driven mode is, therefore,
\be\label{eq:gambeam}
50 {V_{\rm ph}\over\Gamma_*} \la \RNS,
\ee
or equivalently
\be
\bar\gamma \la \left(10^{-2}{\omega_P \RNS\over c}\right)^{2/3}
= 7\times 10^2\, B_{\rm NS,14}^{1/3}\,R_{\rm NS,6}^{2/3}\,
\left({P\over 6~{\rm s}}\right)^{-1/3}\,\left({J\over\rhoGJ c}\right)^{1/3}.
\ee
Comparing this expression with eq. (\ref{eq:gamcrit}) for the 
equilibrium Lorentz factor $\gamin = 80\,B_{\rm NS,14}
(\kB\Tbb/0.5~{\rm keV})^{-1}$ of the pair-creating charges
(eq. [\ref{eq:gamcrit}]), one sees
that this inequality is easily satisfied.  Note that the
growth of the beam-driven mode is not suppressed by the radial
inhomogeneity of the pair plasma, because
growing modes exist over a relatively broad range of wavenumbers.

A similar inequality can be obtained by supposing that the distribution 
function is flat at small momenta, so that the average of $\gamma^{-3}$
over the distribution function is dominated by a small fraction 
$\sim 1/\bar\gamma$ of particles with momenta $p_\parallel \sim \pm m_ec$,
$\langle \gamma^{-3}\rangle^{1/2} \sim 1/\bar\gamma^{1/2}$.
A transrelativistic particle propagating through 
the background electric field gains energy
at the rate $\gamma^{-1}{d\gamma/dt}\sim \bar\gamma c/\RNS$, which when
equated with the growth rate (\ref{eq:growth}) gives the
same dependence on $\bar\gamma$ as in eq. (\ref{eq:gambeam}).

The rate of pair creation changes in response to the beam-driven mode, 
and screening of $E_\parallel$ by the pairs can suppress growth.  
We imagine that gamma rays are emitted and converted to free pairs, and that
one of the particles so created immediately 
reverses direction in the background electric 
field.  There is a critical value of the total free path for this process, 
denoted by $\Delta z$, below which the screening effect is important,
\be\label{eq:dzmax}
\Delta z \;<\; k_*^{-1} \;\sim\; \bar\gamma^{3/2}\,{c\over \omega_P}
\;=\; 0.5\,{\bar\gamma^{3/2}\over B_{\rm NS,14}^{1/2}}\,
\left({P\over 6~{\rm s}}\right)^{1/2}\qquad {\rm cm}.
\ee
When this inequality is violated, the gamma rays that are emitted within a
localized potential perturbation convert to pairs at distant points in 
the circuit, where they cannot influence the growth of the potential.  
The minimal rate of pair creation needed to screen the growing,
beam-driven potential fluctuation $\delta\Phi$ before a broad
spectrum of electrostatic turbulence develops is
\be
e\delta\dot n_\pm > {1\over 50}\Gamma_* \delta\rho.
\ee
Here 
\be
\delta \rho = {k_*^2\delta\Phi\over 4\pi},
\ee
is the corresponding charge density fluctuation.
The change in the rate of gamma-ray emission per unit volume 
resulting from a potential fluctuation $\delta\Phi$ is given by
\be
\delta \dot n_\gamma = {\partial \dot n_\gamma\over\partial\bar\gamma}
{e\delta\Phi\over m_ec^2}.
\ee
If inequality (\ref{eq:dzmax}) is satsified, then $\delta\dot n_\pm \sim \delta
\dot n_\gamma$, and we have a second condition for effective screening
\be\label{eq:kparmin}
k_* <  \left(10^2\,{4\pi e^2\over m_ec^3}{\partial \dot n_\pm\over
\partial\bar\gamma}\right)^{1/3}.
\ee
The background rate of pair creation is given by eq. (\ref{eq:scatrate}), 
namely $e\dot n_\pm \sim (c/2)d\rhoGJ/dr = (Ac/2\RNS)\rhoGJ$.
Substituting for $k_*$ into eq. (\ref{eq:kparmin}), and neglecting 
the difference between $J$ and $\rhoGJ c$, one finds
\be\label{eq:gamsc}
\bar\gamma > \left({1\over 10^2 A}{\omega_P\RNS\over c}\right)^{2/7}
\left({\partial\ln\dot n_\pm\over\partial\ln\bar\gamma}\right)^{-2/7}
= 20\, A^{-2/7}B_{\rm NS,14}^{1/7}\,R_{\rm NS,6}^{1/7}\,
\left({P\over 6~{\rm s}}\right)^{-1/7}\,
\left({\partial\ln\dot n_\pm\over\partial\ln\bar\gamma}\right)^{-2/7}.
\ee

The screening of the growing potential perturbation
requires that both conditions (\ref{eq:dzmax}) and 
(\ref{eq:gamsc}) be satisfied.  The second condition is generally
satisfied by our circuit model.  The first condition is easily
satisfied if $\BNS > 4\BQ$, so that pair creation follows resonant
scattering;  but it is not satisfied if $B < 4\BQ$.  Then
the two-stream instability in the pair plasma will be effectively
triggered near the neutron star surface, and the non-linear growth
of the potential fluctuation into a strong diode may result.

\section{Discussion}\label{s:ten}

We now confront the circuit physics described in this paper
with a few key observational tests, and then summarize
our conclusions.   The behavior of the radio magnetars
appears to be unusual in several respects in comparison with long-period
radio pulsars.

\subsection{Application to the Radio Magnetars XTE J1810$-$197
and 1E 1547.0$-$5408}\label{s:radmag}

\subsubsection{Radio Luminosity}\label{s:radlum}

The spindown of XTE J1810$-$197 is transiently accelerated
(Camilo et al. 2007a).  The polar magnetic field that
is deduced from the standard magnetic dipole formula has
decreased from $B_{\rm pole} = 
2B_{\rm MDR} \sim 5\times 10^{14}$ G to $3\times 10^{14}$ G,
and a further decrease is possible.
The polar field of 1E 1547.0$-$5408 is inferred to be 
$B_{\rm pole} \sim 2\times 10^{14}$ G.  As a result, the 
actual polar fields of these neutron stars could be weaker
than $4\BQ$, in which case the gap model described in 
\S \ref{s:four} is relevant.   We first consider whether
the power dissipated in the gap can supply the observed
radio output.

When the detected pulsed radio emission of XTE J1810$-$197
was at its peak, the Goldreich-Julian charge flow on the open
field lines was $I_{\rm GJ}/e \simeq 
B_{\rm MDR} R_{\rm NS}^3 \Omega^2/ec = 2\times 10^{31}$ s$^{-1}$. 
The mean energy per outflowing particle crossing the gap
is $e\Phi_{\rm gap} = \gamma_{\rm max} m_ec^2 \sim 5\times 10^{-3}$ erg
(eq. [\ref{eq:gmax}]), including a factor $\sim 0.5$ reduction
in the gap voltage due to the screening effects of secondary modes
of pair creation at a temperature $\kB\Tbb \sim 0.25$ keV
(Fig. \ref{f:voltage}).  The net dissipated power is 
therefore $I_{\rm GJ}\Phi_{\rm gap} = 1\times 10^{29}$ ergs s$^{-1}$.
The peak isotropic output at 40 GHz is, by contrast 
$4\pi D^2 \nu S_\nu = 1\times 10^{30}$ ergs s$^{-1}$ at $\nu = 
100$ GHz at a distance $D = 3$ kpc (Camilo et al. 2007a). 
The apparent radio output could be enhanced by an order
of magnitude by the effects of beaming, but even in that 
case the two results are barely consistent.  

We must therefore consider two alternative possibilities:  that the same
gap-like structure can form on a broader bundle of magnetic field lines,
extending into the closed magnetosphere;  or that the power dissipated
on the open field lines is much higher than
is implied by the gap model of \S \ref{s:four}.  It should be kept in
mind that the spindown power of XTE J1810$-$197 was at least
$2.4\times 10^{33}\,I_{\rm NS,45}$ ergs s$^{-1}$ at its peak, and so
it can easily accommodate the radio and infrared output ($L_{\rm IR}
\simeq 1\times 10^{31}$ ergs s$^{-1}$: Rea et al. 2004; Wang et al. 
2007).

\subsubsection{Peak Emission Frequency and Pulse Width}\label{s:peakem}

The pulse duty cycle of the two radio magnetars is much larger than would
be expected if the radio emission originated close
to the star on the open magnetic field lines.  
We deduce either that the emission zone is close to the light cylinder,
or that there is coherent emission from the closed magnetic field lines.
In the case of
1E 1547.0$-$5408, this effect could be partly explained by a nearly
alignment of the magnetic axis with the rotation axis (Camilo et al. 2008).
Nonetheless, the fit to the linear polarization
curve found by these authors does not require that the tilt angle
between $\bmu$ and $\Om$ be nearly as small as $\theta_{\rm open}(\RNS)$,
if the radio beam is intrinsically broad.  (Indeed, the small apriori 
probability of detecting a radio beam of angular width 
$\sim 2\theta_{\rm open}(\RNS)$ from this object can be used to argue 
that the beam must be broad.)

It is possible to deduce a value of $\gamma\Mpm$ under the assumptions
that i) the radio waves are emitted tangent to the magnetic field lines
within a maximum angle which is a multiple $N_\theta$ of $\theta_{\rm open}(r)$;
and ii) the emission frequency at radius $r$ is equal to a multiple
$N_\nu$ of $\nu_{P\pm}^{\rm rel}$.   
We consider an outflowing gas of pairs, within which the plasma frequency is
\be
\nu_{P\pm}'
= {\omega_{P\pm}'\over 2\pi} = \left({2n_\pm' e^2\over \pi m_e}\right)^{1/2}
\ee
in the bulk frame.  The proper density of electrons and positrons is
\be\label{eq:n2pair}
2n_\pm' = {1\over \gamma}\left(n_{e+} + n_{e^-}\right) 
= {\Mpm \over\gamma} n_{\rm GJ}
\ee
where $\Mpm \gg 1$ is the number of light charges created 
per primary Goldreich-Julian charge.   We assume for simplicity
that the plasma is in bulk motion with a Lorentz factor $\gamma$, so that
the frequency is increased in the stellar frame by a factor $2\gamma$,
\be\label{eq:omp1}
\nu_{P\pm}^{\rm rel} = \left({8n_\pm e^2\over \pi m_e}\gamma\right)^{1/2}.
\ee
The plasma frequency is maximized close to the surface
of the star,
\ba\label{eq:numax}
\nu_{P\pm}^{\rm rel}(\RNS) &=&
\left[\gamma\Mpm {2\Omega e\BNS
\over \pi^2 m_ec}\right]^{1/2}\nn
&=&
2.6\times 10^{12}
\,\left({\gamma \Mpm\over 10^4}\right)^{1/2}\,
\left({\BNS\over 4\BQ}\right)^{1/2}
\left({P\over 6~{\rm s}}\right)^{-1/2}\qquad {\rm Hz}.
\ea
Outside the maximum radius of pair creation $R_\pm$,
one obtains the usual scaling,
\be\label{eq:nupe}
\nu_{P\pm}^{\rm rel}(r) =
\left({r\over R_\pm}\right)^{-3/2}\;
\nu_{P\pm}^{\rm rel}(R_\pm)\qquad (r > R_\pm).
\ee
Note that $\nu_{P\pm}^{\rm rel}$ depends on the product $\gamma\Mpm$,
which is approximately invariant if the charges undergo resonant
scattering and the scattered photons are able to re-convert to pairs.
In the gap model described in \S \ref{s:four}, one has
$\gamma\Mpm \simeq e\Phi_{\rm gap}/m_ec^2$.

The tangent to a dipole field lines makes an angle ${3\over 2}\theta$
with respect to the magnetic axis when $\theta \ll 1$, and so the
pulse duty cycle $\delta t/P$ is given by
\be
{\delta t\over P} = {2N_\theta \left({3\over 2}\theta_{\rm open}\right)\over 2\pi}
= N_\theta{3\theta_{\rm open}\over 2\pi}.
\ee
The multiplicity corresponding to a duty cycle $\delta t/P$ at a
frequency $\nu = N_\nu\nu_{\rm P\pm}^{\rm rel}$ is
\be
\gamma\Mpm = 
{3\times 10^6\over N_\nu^2N_\theta^6}\,\nu_{\rm GHz}^2\,
\left({\delta t/P\over 0.1}\right)^6\,\left({P\over 6~{\rm s}}\right)^4
\left({\BNS\over 4\BQ}\right)^{-1}.
\ee
Substituting numbers appropriate to XTE J1810$-$197 ($\delta t/P = 0.2$
at $\nu = 2$ GHz and $P = 5.44$ s) gives
\be\label{eq:gmtot}
\gamma\Mpm \sim 5\times 10^8\,N_\nu^{-2}N_\theta^{-6}\qquad\qquad
({\rm XTE~J1810-197}).
\ee

Let us first compare this result with the turbulent flux tube model 
that is described in Thompson (2008).
In that model, the pair multiplicity is $\gamma\Mpm \sim (eI_{\rm GJ}/m_ec^3) = 
1\times 10^8$.  The plasma frequency at the base of the flux tube then sits
in the optical-IR band, and it is possible for the optical-IR
emission of XTE J1810$-$197 to be an upward extension of the
spectrum at $\sim 100$ GHz.  If this were the case, a narrower pulse profile would 
be expected in the IR band as compared with the radio band.  

The alternative explanation for the broad pulses and high intensity
of the radio emission of XTE J1810$-$197 is that the outer closed
magnetic field lines support a unidirectional discharge.
Within a critical angle $\theta(\RNS) \simeq 3^\circ\,(L_X/10^{34}~
{\rm ergs~s^{-1}})^{-1/2} \sim 9\theta_{\rm open}(\RNS)$ from the magnetic
axis, outflowing relativistic charges are not
stopped by resonant scattering of the thermal X-rays.  The 
gap model of \S \ref{s:four} gives $\gamma\Mpm \sim 1\times 10^4$,
which is consistent with eq. (\ref{eq:gmtot}) 
if $N_\nu^{1/2}N_\theta \sim 6$.  The observed radio power is also
reproduced, because the cross-sectional area of the emitting field
lines is $30-100$ times larger.

Although we cannot in this paper decide between these two possibilities,
the spectrum of the radio magnetars provides an important clue.
The very high peak frequency of the emission does not, by itself, demand 
a high plasma density on the open magnetic field lines.  But the comparison
with ordinary radio pulsars (for which the peak frequency is $\sim 0.1-1$ GHz)
does suggest that the plasma density is much higher in the radio magnetars
than in pulsars, by a factor $> 10^4-10^6$ if one uses the scaling
$n_\pm \propto \nu_{\rm peak}^2$.

\subsection{Is $4\BQ = 1.8\times 10^{14} G$ a Limiting Polar Magnetic
Field for Radio Pulsars?}

The distribution of neutron star magnetic fields, as inferred from pulsar spindown,
offers important tests of the physics
explored in this paper.   Considering only sources that are discovered in the
radio band,\footnote{As noted above,
the polar fields of the radio magnetars cannot be deduced with precision from
their spindown rates due to the presence of large-amplitude torque variations,
but are not obviously inconsistent with this bound.} 
the distribution of dipole fields has an upper envelope at
$B_{\rm MDR} \simeq 1\times 10^{14}$ G.  The corresponding polar magnetic field
is remarkably close to $4\BQ = 1.8\times 10^{14}$ G.  A previous suggestion
that radio emission could be quenched in super-QED magnetic fields 
(Baring \& Harding 1998) depends on assumptions about the polarization
dependence of photon splitting that are not born out by microphysical
calculations (Adler 1971; Usov 2002).  Our focus here is different,
on the voltage of the open-field circuit and the properties of the gap
that may form there.

When the surface magnetic field is stronger than $4\BQ$, the open-field voltage
depends on whether a gap is able 
to form displaced from the surface of the star.  The most plausible mechanism of gap formation
involves the generation of multiple weak double layers in the circuit, 
seeded by a two-stream instability, which then accumulate into a diode
structure with a larger voltage.  The two-stream instability is stronger
in the low-voltage circuit solution described in
\S \ref{s:five} than it is the higher-voltage solutions with a gap,
but the absence of a sharp boundary at the surface $B = 4\BQ$ may
impede the second stage of the process.  

Whether or not a gap does form in this situation, pairs are easily created by resonant scattering 
throughout the inner circuit.  Given that the net difference in circuit voltage between the two
solutions is only a factor of 10, the detectability of radio pulsations
from quiescent neutron stars with $B_{\rm NS} > 4\BQ$ and $P \sim 1-10$ s
is probably a matter of flux sensitivity, rather than the presence or absence
of emission.

\subsection{Summary}

\noindent 1. -- We have described a self-consistent circuit solution
with a gap at the surface of the neutron star, which requires
a current density $|J| > \rhoGJ c$ and a surface magnetic field 
$\BNS < 4\BQ$.   The voltage across the gap is
$e|\Phi_{\rm gap}| \sim 10^4 m_ec^2$ when $\kB\Tbb \la 0.15-0.2$ keV.  
We have reconsidered the
effects of magnetic field line curvature on the voltage, and find that
the octopole component makes the largest contribution to linear order,
with the quadrupole contributing to second order,
when these two components are treated as a perturbation to the dipole.
It is not known whether the two detected radio magnetars 
have polar magnetic fields stronger or weaker than the critical
value $4\BQ$, after allowing for changes in magnetospheric structure
resulting from the injection of currents during periods of X-ray activity. 

\vskip .1in
\noindent 2. -- The dominant pair creation mechanism within the gap
is the conversion of resonantly scattered photons off the magnetic field
when the blackbody temperature is relatively low, $\kB\Tbb \la 0.2$ keV. 
At higher temperatures, the voltage is reduced by secondary modes
of pair creation.  Up to a temperature of $\sim 0.3$ keV, the most 
important such mechanism involves the non-resonant
(Klein-Nishina) scattering of X-rays, followed by direct conversion 
off the magnetic field.  At yet higher temperatures, or in fields
weaker than $\sim 10^{14}$ G, collisions between gamma
rays and thermal X-rays occur at a higher rate.  The net effect is
to reduce the gap voltage by a factor $\sim 0.1$ when $\kB\Tbb \sim 0.3$ keV.

\vskip .1in
\noindent 3. -- Screening of the electric field outside the gap can
be effected by the polarization of the outflowing pairs.  The pairs
injected near the outer boundary of the gap continue to downscatter
off the ambient thermal X-rays, thereby creating secondary pairs.
We have deduced the spectrum of energies that result.  Inside
a distance $r \sim 10\,\RNS$, the outflowing pair cloud maintains
a charge density equal to $\rhoGJ$, as charges of the opposite sign
reverse direction and return to the star.  Mildly relativistic electrons
and positrons feel a strong drag force near $r \sim 10\RNS$, but
a steady, charge-balanced
flow remains possible in which the `negative' charges flow
outward sub-relativistically.  

\vskip .1in
\noindent 4. -- Where $B > 4\BQ$, the open-field circuit sustains a 
much lower voltage, because resonantly scattered
photons convert directly to free pairs.  In the temperature
range $\kB\Tbb \sim 0.1-0.5$ keV, the voltage is reduced by a factor
$\sim 0.05-0.1\,(\BNS/4\BQ)$ as compared with the gap model evaluated at 
the threshold field $\BNS = 4\BQ$.  We have also derived the voltage
distribution through the inner circuit in the case where pair creation 
is smoothly distributed,
at just the rate that is needed to compensate the gradient in $\rhoGJ$.  
There may be a connection between the quenching of a surface gap 
at $\BNS > 4\BQ$ and the absence of radio pulsars with dipole fields
stronger than $10^{14}$ G.


\vskip .1in
\noindent 5. -- The radio output of XTE J1810$-$197 and 1E 1547.0$-$5408 
is somewhat brighter than is allowed by the gap model of  \S \ref{s:four},
if the emission is restricted to the open dipolar field lines.  
The radio pulses are also much broader than is typical of long-period
radio pulsars (Camilo et al. 2006, 2007b).  The same
gap model predicts much narrower pulses than are observed
if the radio emission frequency is close to the local plasma frequency,
and the emission comes from the open magnetic field lines.  When combined
with the very hard radio spectra, this suggests that the rate of pair creation
is much higher in the radio magnetars than in more typical
radio pulsars.  The wide pulses, hard spectra, and rapid variability
of the radio magnetars are more consistent with the electrodynamic
model described in Thompson (2008), in which a broad spectrum of
Alfv\'enic turbulence is excited by current-driven instabilities
in the outer magnetosphere.

\vskip .1in
\noindent 6. -- Relativistic pairs flowing outward from an
X-ray bright magnetar will suffer strong resonant drag 
beyond a radius $R_{\rm col} \sim 50\,(P/6~{\rm s})^{1/3}R_{\rm NS}$
if the guiding field line is displaced too far from the magnetic axis.
At a spin period of $6$ seconds, this critical
angle is $\sim 10\,\theta_{\rm open}$; but if
the spin is faster than $\sim 0.4\,(L_X/10^{34}~{\rm ergs~s^{-1}})^{3/4}$ s,
then most of the secondary pairs on the open field lines will suffer 
strong drag outside $\sim 10$ stellar radii.   Radio emission from
these charges will be suppressed by induced scattering unless they
are continuously reaccelerated.  There is therefore
a bias against detecting
pulsed radio emission from neutron stars with moderately fast
spins and high X-ray luminosities (e.g. young magnetars).

\acknowledgments

This work was supported by the NSERC of Canada.  It began
as an investigation of the radio death line of magnetars, and I am
indebted to Andrei Beloborodov for his contribution to those early
efforts.  In particular, he provided a cogent argument that
pair cascades driven by curvature radiation cannot be self-consistently
sustained in magnetar magnetospheres.  I also thank him for suggesting
some simplifications in the presentation of \S \ref{s:nogap},
and a correction to the calculation of resonant drag in \S \ref{s:thetmax}.
Part of this work was carried out during a visit to Cornell
University, which I thank for its hospitality.

\begin{appendix}

\section{Sources of Gradient in Corotation Charge Density}\label{s:GR}

The voltage on the open magnetic field lines is dominated by the
gradient in the corotation charge density in the case where the
first order difference between $J$ and $\rhoGJ c$ is compensated
by a returning flow of charges of the sign opposite to the local
corotation charge.  Here we compare the effect of general relativistic
frame dragging (Muslimov \& Tsygan 1992) with field line curvature
as dominated by a quadrupolar and octopolar magnetic field.

A purely dipolar magnetic field in a Schwarzschild spacetime has the shape
\be
 {\bf B}(r) = \frac{f(r)}{f(\RNS)}\,{\bf B}_{\rm dipole}(r),
\ee
where $f(r)=1+\Sigma_{k=0}^\infty \frac{3}{k+3}(2G\MNS/c^2r)^k$
(Muslimov \& Tsygan 1992, eq.~32).
The prefactor here describes the deviation
of the poloidal magnetic field lines from the shape they would
have in the absence of gravity.  It does not affect the voltage
directly, because both the charged particle flux and the magnetic
flux satisfy the same conservation equation.   The relativistic
charge density $\alpha(r)\rhoGJ(r)$ has the form (\ref{eq:rhogjrelc}),
taking into account the effects of frame dragging.   
When evaluating the lapse function, the stellar rotation can be
neglected:  on has
\be
  \alpha(r) = \left(1-{2GM_{\rm NS}\over c^2 r}\right)^{1/2}
\ee
outside the star.  It is sufficient to take $\alpha \rightarrow 1$ 
and $f \rightarrow 1$ in all expressions, because the magnetic flux
density and
the relativistic current density ${\bf J} = \alpha \rho {\bf v}$ and the
magnetic flux density satisfy the same conservation law\footnote{Here
$\bnabla$ is the gradient operator in Minkowski space.}
$\bnabla\cdot\B = \bnabla\cdot{\bf J} = 0$.

   An additional mechanism for voltage generation becomes 
   significant for a more general magnetic configuration
   in which the open field lines are strongly curved.  The ratio
     $\rhoGJ/B$ increases with distance from the star on those field 
     lines that bend toward the spin axis. In the absence of pair creation
     this implies an increasing mismatch $|\rho| > |\rho_{\rm GJ,0}|$, 
   and a voltage 
     is generated that accelerates the Goldreich-Julian flow 
     (e.g. Barnard \& Arons 1982; Asseo \& Khechinashvili 2002).
     This effect dominates over the frame-dragging effect beyond a certain
     radius, and can be especially important in magnetars due to the
     screening of the parallel electric field by direct pair creation close
     to the star.  

We focus here on the case where the magnetic field is predominantly
dipolar beyond zone I.  The presence of a higher multipole (quadrupole
or octopole) causes a bending of the open dipolar field lines.  The
angular deflection $\delta\theta(r)$ 
can be deduced by working inward from the light
cylinder.  The radial component of the magnetic field is dominated
by the dipole component, $B_D = B_{\rm NS} (r/\RNS)^{-3}$.  
Well inside the light cylinder, but still far from the magnetar surface,
the non-radial field component on the open-flux bundle is dominated 
by the higher multipole, e.g.,
\be\label{eq:bnr}
B_\theta(r) \simeq B_\ell(\RNS)\left({r\over \RNS}\right)^{-\ell-2}
\equiv \varepsilon_\ell\,B_{\rm NS}\,
\left({r\over \RNS}\right)^{-\ell-2}.
\ee
We suppose that this additional multipole is oriented randomly
with respect to the magnetic dipole axis.  Then
\be
\delta\theta(r) \simeq \int_r^\infty {B_\theta(r)\over B_D(r)}\,{dr\over r}
= {\varepsilon_\ell\over \ell-1}\,\left({r\over \RNS}\right)^{-\ell+1}.
\ee
The parameter $\varepsilon_\ell$ contains the angular dependence of
the $\ell$-pole.  

The lowest-order multipole component which makes a significant contribution
to voltage and curvature of the open field lines is the octopole.\footnote{To
linear order in the multipole amplitude.  This
conclusion presupposes the dominance of the dipole component of
the magnetic field in the acceleration zone.  If the dipole component
were absent, then a quadrupole by itself would still not supply significant
field line curvature.  The opening angle in that case would be
$\theta_{\rm open} \sim \Omega r/c$, and the curvature radius
would be comparable to the light cylinder radius,
$R_C \sim r/\theta_{\rm open} \sim c/\Omega$.  A mixture of multipoles
is required to provide an effective particle accelerator.}
To see this, note that the transverse deflection of the open magnetic flux
bundle away from a pure dipole is $L_\perp = r\delta\theta(r) \propto
r^{2-\ell}$, which is constant for a quadrupole.  More generally
the curvature radius can be found from
\be
R_C^{-1} = \left|(\hat B\cdot\bnabla)\hat B\right|,
\ee
which in combination with 
\be
\hat B \simeq
\left[1-b_\theta(r)^2\right]^{1/2}\hat r + b_\theta(r)\hat \theta;
\qquad
b_\theta(r) \equiv {B_\ell(r)\over B_D(r)}
\ee
gives
\be\label{eq:rcfull}
R_C^{-1} =
{1\over r}\left|\left({\partial(r b_\theta)\over \partial r} + 
 b_\theta{\partial b_\theta\over \partial\theta}\right)\hat\theta 
   - b_\theta{\partial(rb_\theta)\over\partial r}\hat r\right|.
\ee
to second order in $b_\theta$.
A single additional multipole imparts a curvature radius
\be
R_C = {r\over (\ell-2)(\ell-1)\delta\theta(r)} = 
{\RNS\over (\ell-2)\varepsilon_\ell}\,
\left({r\over \RNS}\right)^\ell,
\ee
to lowest order in $b_\theta$.
This expression gives infinite $R_C$ for $\ell = 2$, and
\be\label{eq:rcnonlin}
R_C = {r\over 2\delta\theta} = 
\left(\varepsilon_2{\partial\varepsilon_2\over\partial\theta} + 
\varepsilon_3\right)^{-1}\,\left({r\over \RNS}\right)^3\,\RNS
\ee
when an octopole is added to the dipole.  The quadrupole provides
a second-order correction to the octopole curvature, which has the
same scaling in radius, and is of comparable magnitude if
$\varepsilon_3 \sim \varepsilon_2^2$.

When the magnetic field is mainly dipolar, the opening angle
is given approximately by the dipole formula,
\ba\label{eq:thetop}
\theta_{\rm open}(r) &=& 
\left[f_{\rm open}(\chi){\Omega r\over c}\right]^{1/2}\nn
 &=&   0.014\,f_{\rm open}^{1/2}\,R_{\rm NS,6}^{1/2}\,
       \left({P\over {\rm s}}\right)^{-1/2}\,
    \left({r\over \RNS}\right)^{1/2}\;\;\;\;{\rm rad}.
\ea
The factor $f_{\rm open}(\chi) = O(1)$ corrects for an inclination angle
$\chi \neq 0$ between the magnetic moment and the rotation axis; 
we generally set $f_{\rm open} = 1$ in expressions in this paper.
The tilt of the field lines with respect to the magnetic dipole
axis can easily be much larger than $\theta_{\rm open}$.  For example, a
quadrupole/octopole component of magnitude
\be\label{eq:epdef}
\eps3 \equiv
\varepsilon_3 + \varepsilon_2{\partial\varepsilon_2\over\partial\theta}
> 0.012\,f_{\rm open}^{1/2}R_{\rm NS,6}^{1/2}\,
\left({P\over 6~{\rm s}}\right)^{-1/2}\,\left({r\over \RNS}\right)^{1/2}
\ee
is sufficient to impart a tilt 
$\delta\theta > \theta_{\rm open}$ at a radius $r$.
In this situation, both $\delta\theta$ and $\rho_{\rm GJ}$ are approximately
constant across the open flux bundle.

The electrostatic potential, as given by eq. (\ref{eq:potparb}),
depends on the deviation of the charge density from the local corotation
density.\footnote{For simplicity we
set the lapse function $\alpha \rightarrow 1$ in what follows.}
In this case,
there is a direct relation between the curvature of the open flux bundle
and the gradient in $\rho_{\rm GJ}$ parallel to the magnetic 
field.  One has
\be\label{eq:drhogjdl}
{d(\rho_{\rm GJ}/B)\over dl}
= {\Om\over 2\pi c}\cdot\left[ (\hat B\cdot\bnabla)\hat B\right]
  = {\Omega \over 2\pi c R_C}\,\sin\chi\cos\phi_C,
\ee
or, equivalently,
\be
B{d(\rho_{\rm GJ}/B)\over dr} = {\rho_{\rm GJ}\over R_C}\,
\sin\chi\,\cos\phi_C.
\ee
We have made use of the fact that $(\hat B\cdot\bnabla)\hat B$ points
in the $\theta$-direction in the plane of curvature of the open magnetic
field lines.\footnote{To leading order in the quadrupole/octopole amplitudes;
eq. [\ref{eq:rcfull}].}
Here $\phi_C$ is the angle by which the plane of curvature is
rotated with respect to the $\bmu-{\bf\Omega}$ plane.  The zero of
$\phi_C$ is defined so that the field lines bend {\it toward} the
rotation axis when $\cos\phi_C > 0$.  There is no acceleration when
$\cos\phi_C < 0$. 

\subsection{Limiting Voltage on a Slender Magnetic Flux Tube.}

For completeness, we examine the voltage in an extended gap which
has a length $\Delta r$ much larger than the radius $\theta_{\rm open} r$
of the open magnetic field bundle.
When the star rotates steadily and the magnetic field is fixed in the 
corotating frame, an electrostatic potential $\Phi(r)$ can be introduced,
which satisfies the equation,
\be
   \bnabla\cdot\left(\alpha^{-1}\bnabla\Phi\right)=-4\pi(\rho-\rhoGJ).
\ee
Then the electric field component along $\bB$ is given by 
\be
  E_\parallel= \alpha^{-1}\hat B\cdot\bnabla\Phi.
\ee
Once again, we make the approximation $\alpha \rightarrow 1$.
We approximate the open flux bundle as being locally cylindrical,
with a cross-sectional area
\be\label{eq:Sperp}
S_\perp \simeq \pi \theta_{\rm open}^2 r^2.
\ee
Choosing  a longitudinal coordinate $r$ and transverse (cylindrical) radius
$\varpi = \theta r$,
Gauss' law can be expressed as
\be\label{eq:gaussc}
{1\over\varpi}{\partial\over\partial\varpi}\left[
  \varpi {\partial \Phi\over\partial\varpi}\right]
  + {d^2\Phi\over dr^2} = -4\pi(\rho-\rho_{\rm GJ}).
\ee

It is usual to apply the boundary condition $\Phi = 0$ at
$\varpi = \theta_{\rm open} r$, 
which must however be reconsidered
when the closed magnetic field lines 
are twisted by the release of internal stresses.  
In this situation, a modest voltage $\sim 10^2-10^3$ V develops on the
closed field lines, as is required to sustains the current
(Beloborodov \& Thompson 2007).  
The power dissipated in the closed-field circuit is drawn from
the energy of the toroidal magnetic field, and the gradual relaxation of
this twist is associated with a time derivative
of the longitudinal component of the vector potential.
The twist on the open magnetic field lines is, by contrast,
continually replenished by the rotation of the star, and
decreases only adiabatically as the star spins down.
We therefore retain the boundary condition
$\Phi = 0$ at $\varpi = \theta_{\rm open} r$, but it should
be understood that this boundary condition could be modified
if the outer magnetosphere were subject to fast, current-driven
instabilities.

A simple solution to eq. (\ref{eq:gaussc}) is obtained by neglecting
the longitudinal component of the laplacian and taking both $\rho$
and $\rho_{\rm GJ}$ to be constant across the open flux bundle.
This approximation is reasonably accurate outside a distance 
$r-\RNS \sim (\theta_{\rm open}/2) \RNS$ from the neutron
star surface. The solution is
\be\label{eq:potparb}
\Phi(r,\theta) 
= {\IGJ\over c}\,\left({\rho\over\rho_{\rm GJ}}-1\right)
\,\left(1 - {\theta^2\over\theta_{\rm open}^2}\right).
\ee
Here
\be
  \IGJ(r) = S_\perp(r) \alpha(r)\rho_{\rm GJ}(r) c
\ee
is the net current along the open field-line bundle.
The twist on the open magnetic field lines is supported by
the outward flow of the corotation charge.
Neglecting the effect of frame dragging, the charge flow rate is given by
\be\label{eq:igj}
   {|\IGJ^0|\over e} = {\Omega^2 B_{\rm NS} \RNS^3\over 2ec}
   = 4\times 10^{31}\,B_{\rm NS,15}R_{\rm NS,6}^3\,
  \left({P\over 6~{\rm s}}\right)^{-2}\;\;\;\;{\rm s}^{-1}.
\ee

An important feature of eq. (\ref{eq:potparb}) is the reversal in sign
compared with the solution that would be obtained by neglecting
the transverse component of the laplacian in favor of the
longitudinal component, and applying the boundary condition
$\Phi(\RNS,\theta) = 0$. This means that if the charge
density $\rho$ decreases with radius more rapidly than $\rho_{\rm GJ}$,
then an electric field is generated that is of the right sign to
drive the charges away from the star (Scharlemann et al. 1978; 
Muslimov \& Tsygan 1992).  

We suppose that screening is absent beyond some radius $R_*$.
Enough charges are created inside this radius
so that $\rho = \rho_{\rm GJ}$; but that the charge density
is determined by the continuity of the outward charge flow
farther out in the magnetosphere.  The voltage solution beyond
this radius is, then
\be\label{eq:potparc}
  \Phi(r,\theta) = \kappa_*\,{\IGJ^0\over c}\,
  \left[\left({r\over R_*}\right)^{-3}-1\right]\,
  \,\left(1 - {\theta^2\over\theta_{\rm open}^2}\right).
\ee
Here we have used
\be
\kappa_* \equiv \kappa(R_*) = {2G\INS\over R_*^3 c^2}.
\ee
Expanding equation (\ref{eq:potparc}) near $R_*$ gives
\be\label{eq:potpard}
   \Phi(R_*+\Delta r) - \Phi(R_*) \simeq - 3 \kappa_*
 {\IGJ^0\over c}\left({\Delta r\over R_*}\right).
\ee
The limiting Lorentz factor
of an electron (or positron, if $\rhoGJ>0$) flowing beyond $r = R_*$ is
\ba\label{eq:gamrel}
   \gmax(R_*)
 &=& {e[\Phi(r\gg R_*)-\Phi(R_*)]\over m_ec^2}  =
   \kappa(R_*)\,{e\IGJ^0\over m_ec^3} \nn
   &=&  {G I_{\rm NS} \Omega^2 eB(R_*)\over m_ec^6}
   = 5.3\times 10^7\,I_{\rm NS,45}\,
    \left[{B(R_*)\over 10^{15}~{\rm G}}\right]\,
    \left({P\over 6~{\rm s}}\right)^{-2}.
\ea

Now let us turn to the case where the gradient in $\rhoGJ$ is dominated
the quadrupole and octopole magnetic field.
The voltage that develops in the absence of screening can be obtained by
integrating eq. (\ref{eq:drhogjdl}) along ${\bf B}$,
\be\label{eq:phimult}
\Phi(r) =  \kappa_\ell\,\frac{\Omega^2\mu}{(\ell-1)c^2}\,
    \left({R_*\over \RNS}\right)^{-\ell+1}
    \left[1-\left({r\over R_*}\right)^{-\ell+1}\right] \qquad r>R_*,
     \;\; \ell \geq 3.
\ee
In this expression,
$\mu={1\over 2}\BNS\RNS^3$ is the magnetic moment of the star, and
\be\label{eq:Kdef}
\kappa_\ell \equiv (\ell-2)\varepsilon_\ell\,\sin\chi\,\cos\phi_C.
\ee
The expansion near $r = R_*$ is given by
\be\label{eq:phimultb}
 \Phi(R_*+\Delta r) - \Phi(R_*) \simeq  \kappa_\ell\,{I_{\rm GJ}^0\over c}
    \left({R_*\over \RNS}\right)^{-\ell+1}
    \left({\Delta r\over R_*}\right).
\ee
The limiting
Lorentz factor of accelerated electrons (or positrons, if $\rhoGJ>0$) 
that can be accelerated outside a radius $R_*$ is
\be\label{eq:gamrelb}
   \gmax(R_*) = {\kappa_\ell\over\ell-1}{eI_{\rm GJ}^0\over m_ec^3}
   \left({R_*\over \RNS}\right)^{-\ell+1}.
\ee

We can now compare the voltages that are induced by the frame dragging
and field-line curvature effects.  Outside a radius $R_* > \RNS$, 
one has
\be\label{eq:phicomp}
{\Phi({\rm field~curvature})\over\Phi({\rm frame~dragging})}
= {\kappa_\ell \over (\ell-1)\kappa(\RNS)}
\left({R_*\over \RNS}\right)^{-\ell+4}
= 6.7\,{\kappa_\ell\over \ell-1}\,{R_{\rm NS,6}^3\over I_{\rm NS,45}}\,
\left({R_*\over \RNS}\right)^{-\ell+4}.
\ee
In the case where quadrupole/octopole components are present, one
can define an effective octopole amplitude
\be
\K3 = \left(\varepsilon_3 + \varepsilon_2{\partial\varepsilon_2\over
   \partial\theta}\right)\,\sin\chi\,\cos\phi_C,
\ee
following eq. (\ref{eq:rcnonlin}).  One then has
\be\label{eq:phicompb}
{\Phi({\rm field~curvature})\over\Phi({\rm frame~dragging})}
= 3.4\,\K3\,{R_{\rm NS,6}^3\over I_{\rm NS,45}}\,
\left({R_*\over \RNS}\right).
\ee
In the particular case where $B=4\BQ$, the
quadrupole/octopolar components of the magnetic field will
dominate the effects of frame dragging if
$\K3 > 0.17\,B_{\rm NS,15}^{-1/3}I_{\rm NS,45}R_{\rm NS,6}^{-3}$.

A deviation of $\rho$ from $\rhoGJ$ also arises due to the 
flaring of the dipolar field lines, even in the absence
of higher multipoles in the magnetic field (Scharlemann et al. 1978).
This third effect is less important\footnote{We consider 
the simplest case of an aligned rotator in making this estimate.}
than frame dragging if
 $d(B_\theta/B_r)d\ln r \simeq \theta_{\rm open}/4 \ll 3\kappa$.
This inequality translates into an upper bound on 
the surface polar magnetic field, 
\be
B_{\rm NS} < 1.3\times 10^{16}\,\left({P\over {\rm 6~s}}\right)^{3/7}
\;\;\;\;{\rm G}.
\ee
The dipole fields of all magnetars easily satisfy this inequality.

\end{appendix}


\begin{thebibliography}{}


\bibitem[Adler(1971)]{1971AnPhy..67..599A} Adler, S.~L.\ 1971, Annals of 
Physics, 67, 599 

\bibitem[Arons \& Scharlemann(1979)]{1979ApJ...231..854A} Arons, J., \& 
Scharlemann, E.~T.\ 1979, \apj, 231, 854 



\bibitem[Asseo \& Khechinashvili(2002)]{2002MNRAS.334..743A} Asseo, E., \& 
Khechinashvili, D.\ 2002, \mnras, 334, 743 

\bibitem[Baring \& Harding(1998)]{1998ApJ...507L..55B} Baring, M.~G., \& 
Harding, A.~K.\ 1998, \apjl, 507, L55

\bibitem[Baring \& Harding(2001)]{2001ApJ...547..929B} Baring, M.~G., \& 
Harding, A.~K.\ 2001, \apj, 547, 929 

\bibitem[Baring \& Harding(2007)]{2007Ap&SS.308..109B} Baring, M.~G., \& 
Harding, A.~K.\ 2007, \apss, 308, 109 

\bibitem[Barnard \& Arons(1982)]{1982ApJ...254..713B} Barnard, J.~J., \& 
Arons, J.\ 1982, \apj, 254, 713 

\bibitem[Beloborodov(2007)]{2007arXiv0710.0920B} Beloborodov, A.~M.\ 2007, 
ArXiv e-prints, 710, arXiv:0710.0920 

\bibitem[Beloborodov \& Thompson(2007)]{2006astro.ph..2417B} Beloborodov, 
A.~M., \& Thompson, C.\ 2007, \apj, 657, 967

\bibitem[Berestetskii et al.(1982)]{llqed} Berestetskii, V.B.,
Lifshitz, E.M., \& Pitaevskii, L.P. 1982, Oxford: Pergamon Press

\bibitem[Blumenthal \& Gould(1970)]{1970RvMP...42..237B} Blumenthal, G.~R., 
\& Gould, R.~J.\ 1970, Reviews of Modern Physics, 42, 237 

\bibitem[Camilo et al.(2006)]{2006Natur.442..892C} Camilo, F., Ransom, 
S.~M., Halpern, J.~P., Reynolds, J., Helfand, D.~J., Zimmerman, N., \& 
Sarkissian, J.\ 2006, \nat, 442, 892 


\bibitem[Camilo et al.(2007a)]{2007ApJ...663..497C} Camilo, F., et al.\ 
2007a, \apj, 663, 497

\bibitem[Camilo et al.(2007b)]{2007ApJ...666L..93C} Camilo, F., Ransom, 
S.~M., Halpern, J.~P., \& Reynolds, J.\ 2007b, \apjl, 666, L93 

\bibitem[Camilo et al.(2008)]{2008arXiv0802.0494C} Camilo, F., Reynolds, 
J., Johnston, S., Halpern, J.~P., 
\& Ransom, S.~M.\ 2008, ArXiv e-prints, 802, arXiv:0802.0494 



\bibitem[Cheng \& Ruderman(1977)]{1977ApJ...214..598C} Cheng, A.~F., \& 
Ruderman, M.~A.\ 1977, \apj, 214, 598 






\bibitem[Daugherty \& Ventura(1978)]{1978PhRvD..18.1053D} Daugherty, J.~K., 
\& Ventura, J.\ 1978, \prd, 18, 1053 

\bibitem[Daugherty \& Harding(1989)]{1989ApJ...336..861D} Daugherty, J.~K., 
\& Harding, A.~K.\ 1989, \apj, 336, 861 


\bibitem[Dermer(1990)]{1990ApJ...360..197D} Dermer, C.~D.\ 1990, \apj, 360, 
197 

\bibitem[Dieckmann(2005)]{2005PhRvL..94o5001D} Dieckmann, M.~E.\ 2005, 
Physical Review Letters, 94, 155001 

\bibitem[Dieckmann et al.(2006)]{2006NJPh....8..225D} Dieckmann, M.~E., 
Shukla, P.~K., \& Eliasson, B.\ 2006, New Journal of Physics, 8, 225 












\bibitem[Goldreich \& Julian(1969)]{1969ApJ...157..869G} Goldreich, P., \& 
Julian, W.~H.\ 1969, \apj, 157, 869 


\bibitem[Gonthier et al.(2000)]{2000ApJ...540..907G} Gonthier, P.~L., 
Harding, A.~K., Baring, M.~G., Costello, R.~M., \& Mercer, C.~L.\ 2000, 
\apj, 540, 907 








\bibitem[Halpern et al.(2005)]{2005ApJ...632L..29H} Halpern, J.~P., 
Gotthelf, E.~V., Becker, R.~H., Helfand, D.~J., \& White, R.~L.\ 2005, 
\apjl, 632, L29








\bibitem[Hibschman \& Arons(2001a)]{2001ApJ...554..624H} Hibschman, J.~A., 
\& Arons, J.\ 2001a, \apj, 554, 624 

\bibitem[Hibschman \& Arons(2001b)]{2001ApJ...560..871H} Hibschman, J.~A., 
\& Arons, J.\ 2001b, \apj, 560, 871 






\bibitem[Ibrahim et al.(2001)]{2001ApJ...558..237I} Ibrahim, A.~I., et al.\ 
2001, \apj, 558, 237 









\bibitem[Kardashev et al.(1984)]{1984AZh....61.1113K} Kardashev, N.~S., 
Mitrofanov, N.~S., \& Novikov, I.~D.\ 1984, \azh, 61, 1113 




\bibitem[Kuiper et al.(2004)]{2004ApJ...613.1173K} Kuiper, L., Hermsen, W., 
\& Mendez, M.\ 2004, \apj, 613, 1173 

\bibitem[Kuiper et al.(2006)]{2006ApJ...645..556K} Kuiper, L., Hermsen, W., 
den Hartog, P.~R., \& Collmar, W.\ 2006, \apj, 645, 556 






\bibitem[Lyubarskii \& Petrova(1998)]{1998A&A...337..433L} Lyubarskii, 
Y.~E., \& Petrova, S.~A.\ 1998, \aap, 337, 433 

\bibitem[Lyutikov \& Thompson(2005)]{2005ApJ...634.1223L} Lyutikov, M., \& 
Thompson, C.\ 2005, \apj, 634, 1223 




\bibitem[Medin \& Lai(2007)]{2007MNRAS.tmp.1065M} Medin, Z., \& Lai, D.\ 
2007, \mnras, 1065 





\bibitem[Mereghetti et al.(2005)]{2005A&A...433L...9M} Mereghetti, S., 
G{\"o}tz, D., Mirabel, I.~F., \& Hurley, K.\ 2005, \aap, 433, L9 

\bibitem[Mestel et al.(1985)]{1985MNRAS.217..443M} Mestel, L., Robertson, 
J.~A., Wang, Y.-M., \& Westfold, K.~C.\ 1985, \mnras, 217, 443 


\bibitem[Muslimov \& Tsygan(1992)]{1992MNRAS.255...61M} Muslimov, A.~G., \& 
Tsygan, A.~I.\ 1992, \mnras, 255, 61 






\bibitem[Rea et al.(2004)]{2004A&A...425L...5R} Rea, N., et al.\ 2004, 
\aap, 425, L5


\bibitem[Ruderman \& Sutherland(1975)]{1975ApJ...196...51R} Ruderman, 
M.~A., \& Sutherland, P.~G.\ 1975, \apj, 196, 51 



\bibitem[Scharlemann et al.(1978)]{1978ApJ...222..297S} Scharlemann, E.~T., 
Arons, J., \& Fawley, W.~M.\ 1978, \apj, 222, 297 





\bibitem[Shibata(1997)]{1997MNRAS.287..262S} Shibata, S.\ 1997, \mnras, 
287, 262 

\bibitem[Shibata et al.(1998)]{1998MNRAS.295L..53S} Shibata, S., Miyazaki, 
J., \& Takahara, F.\ 1998, \mnras, 295, L53 

\bibitem[Shibata et al.(2002)]{2002MNRAS.336..233S} Shibata, S., Miyazaki, 
J., \& Takahara, F.\ 2002, \mnras, 336, 233 

\bibitem[Silantev \& Iakovlev(1980)]{1980Ap&SS..71...45S} Silantev, N.~A., 
\& Iakovlev, D.~G.\ 1980, \apss, 71, 45 




\bibitem[Sturner(1995)]{1995ApJ...446..292S} Sturner, S.~J.\ 1995, \apj, 
446, 292 

\bibitem[Sturrock(1971)]{1971ApJ...164..529S} Sturrock, P.~A.\ 1971, \apj, 
164, 529



\bibitem[Thompson(2008)]{t08}Thompson, C., \apj, submitted
(arXiv:0802:2571)


\bibitem[Thompson \& Blaes(1998)]{1998PhRvD..57.3219T} Thompson, C., \& 
Blaes, O.\ 1998, \prd, 57, 3219 

\bibitem[Thompson et al.(2000)]{2000ApJ...543..340T} Thompson, C., Duncan, 
R.~C., Woods, P.~M., Kouveliotou, C., Finger, M.~H., \& van Paradijs, J.\ 
2000, \apj, 543, 340


\bibitem[Thompson \& Duncan(2001)]{2001ApJ...561..980T} Thompson, C., \& 
Duncan, R.~C.\ 2001, \apj, 561, 980 

\bibitem[Thompson et al.(2002)]{2002ApJ...574..332T} Thompson, C., 
Lyutikov, M., \& Kulkarni, S.~R.\ 2002, \apj, 574, 332 

\bibitem[Thompson \& Beloborodov(2005)]{thompson05} Thompson, C., \&
Beloborodov, A.~M.\ 2005, \apj, 635, 565

\bibitem[Usov(2002)]{2002ApJ...572L..87U} Usov, V.~V.\ 2002, \apjl, 572, 
L87 

\bibitem[Usov \& Melrose(1996)]{1996ApJ...464..306U} 
Usov, V.~V., \& Melrose, D.~B.\ 1996, \apj, 464, 306 





\bibitem[Wang et al.(2007)]{2007ApJ...665.1292W} Wang, Z., Kaspi, V.~M., \& 
Higdon, S.~J.~U.\ 2007, \apj, 665, 1292 




\bibitem[Woods et al.(2002)]{2002ApJ...576..381W} Woods, P.~M., 
Kouveliotou, C., G{\"o}{\u g}{\"u}{\c s}, E., Finger, M.~H., Swank, J., 
Markwardt, C.~B., Hurley, K., \& van der Klis, M.\ 2002, \apj, 576, 381 



\bibitem[]{woods06} 
Woods, P.~M., \& Thompson, C. 2006, in Compact Stellar X-ray Sources,
eds. W.~H.~G. Lewin and M. van der Klis (Cambridge: Cambridge University Press)
547



\bibitem[Zhang \& Harding(2000)]{2000ApJ...535L..51Z} Zhang, B., \& 
Harding, A.~K.\ 2000, \apjl, 535, L51 

\bibitem[Zhang \& Qiao(1998)]{1998A&A...338...62Z} Zhang, B., \& Qiao, 
G.~J.\ 1998, \aap, 338, 62 



\end{thebibliography}
\end{document}